\documentclass[11pt]{article}
\usepackage{fullpage}
\usepackage[vlined,linesnumbered,ruled]{algorithm2e}
\usepackage[round]{natbib}
\usepackage{times,dsfont,hyperref}
\usepackage{url}
\usepackage{amssymb,amsthm}
\usepackage{subcaption}
\usepackage{upref}
\let\oldnl\nl%
\newcommand{\nonl}{\renewcommand{\nl}{\let\nl\oldnl}}
\usepackage{thmtools,thm-restate}
\declaretheorem{theorem}
\declaretheorem[sibling=theorem]{lemma}

\theoremstyle{definition}

\theoremstyle{definition}

\newtheorem{proposition}[theorem]{Proposition}

\usepackage{mathtools,bm,microtype,booktabs,array,enumitem,float,multirow,threeparttable}
\usepackage[table]{xcolor}
\usepackage[upright=true]{derivative}
\usepackage{physics2}
\usephysicsmodule{ab,ab.braket}

\hypersetup{
    colorlinks,
    linkcolor={red!50!black},
    citecolor={blue!50!black},
    urlcolor={blue!80!black}
}
\definecolor{ourresult}{gray}{0.92}
\let\physicsket\ket
\let\physicsbra\bra
\renewcommand{\ket}[1]{\physicsket|#1>}
\renewcommand{\bra}[1]{\physicsbra<#1|}
\renewcommand{\vec}[1]{\bm{#1}}
\newcommand{\CC}{\mathbb C}
\newcommand{\RR}{\mathbb R}
\newcommand{\E}{\mathbb E}
\newcommand{\ind}{\mathds{1}}
\newcommand{\eps}{\varepsilon}
\newcommand{\norm}[1]{\ab\|#1\|}
\newcommand{\Rrate}[1]{R_{d,r}(#1)}
\newcommand{\PhiCost}[1]{\Phi_d(#1)}
\DeclareMathOperator{\tr}{tr}
\DeclareMathOperator{\rank}{rank}
\DeclareMathOperator{\TV}{TV}
\DeclareMathOperator{\Var}{Var}
\DeclareMathOperator{\KL}{D}
\DeclareMathOperator{\kl}{kl}
\providecommand{\argmax}{\operatorname*{arg\,max}}

\title{Optimal Sample Complexity of Purity Estimation and Purest State Identification with Limited Quantum Memory}
\usepackage{authblk}
\makeatletter
\renewcommand\AB@affilsepx{, \protect\Affilfont}
\makeatother
\author[1]{Zhuohua Li}
\author[2]{Maoli Liu}
\author[1]{Cong Tian}
\author[1]{Shengchao Qin}
\author[2]{John C.S. Lui}
\affil[1]{Xidian University}
\affil[2]{The Chinese University of Hong Kong}
\date{}

\begin{document}
\maketitle

\begin{abstract}
We study the sample complexity of purity estimation with limited quantum memory in the two-copy reset model, where an \(r\)-dimensional quantum register can be retained between two input copies but must be reset after each pair.
Classical memory, adaptive measurement selection, and random stopping are unrestricted.
For dimension \(d=2^n\), register dimension \(r=2^k\), and sufficiently small additive error \(\eps\), we establish the optimal sample complexity
\[
\Theta\ab(\max\ab\{\eps^{-2},\min\ab\{\sqrt d/\eps,d/(r\eps^2)\}\}).
\]
This matches the known upper bound and closes the remaining gap in the dependence on estimation accuracy.
Our key technical contribution is a sharp Fisher information bound for the complete adaptive measurement record, accounting for information accumulated across rounds that share a single hidden Haar direction.
We further study fixed-confidence identification of the purest among \(K\) unknown quantum states under the same memory constraint.
We develop a confidence elimination algorithm that identifies the purest state with high probability and establish lower bounds that match the upper bound up to logarithmic factors.
\end{abstract}

\section{Introduction}\label{sec:introduction}

Estimating the purity of an unknown quantum state is a fundamental task in quantum information, with applications in quantum device characterization and benchmarking~\citep{Brydges2019,elben-2020-cross}.
For a quantum state \(\vec\rho\) on \(\CC^d\), its purity is defined as \(P(\vec\rho)=\tr(\vec\rho^2)\), ranging from \(1/d\) for the maximally mixed state to \(1\) for a pure state.
Unlike full state tomography, purity estimation requires only a scalar property of the state, avoiding the need to reconstruct the entire density matrix~\citep{ekert2002direct,all}.
Its sample complexity, however, depends critically on the measurement capabilities available to the learner.
When joint measurements on two copies of \(\vec{\rho}\) are allowed, the SWAP test estimates purity to additive error \(\eps\) using \(O(\eps^{-2})\) copies, independently of the dimension \(d\)~\citep{ekert2002direct}.
In contrast, when each copy must be measured individually, \citet{all} established an upper bound of \(O(\max\{\eps^{-2},\sqrt d/\eps\})\) using randomized measurements.

To understand the tradeoff between these two extremes, \citet{gong} studied purity estimation with limited quantum memory.
In their model, the learner may retain a quantum register of dimension \(r\) between two consecutive copies, but must discard all quantum information after processing each pair.
Classical information can be retained throughout the experiment and used to select subsequent measurements.
Thus, this model interpolates between measurements of individual copies at \(r=1\) and arbitrary joint measurements on each pair at \(r=d\).
They established an upper bound of
\(O(\max\{\eps^{-2},\min\{\sqrt d/\eps,d/(r\eps^2)\}\})\)
and a lower bound of
\(\Omega(\max\{\eps^{-2},\min\{\sqrt{d/\eps},d/(r\eps^2)\}\})\).
However, a gap remains between the \(\sqrt d/\eps\) and \(\sqrt{d/\eps}\) terms.
This leaves open the optimal joint dependence of sample complexity on dimension \(d\), accuracy \(\eps\), and memory \(r\).
Closing this gap requires a sharp lower bound against arbitrary adaptive measurements, which can exploit earlier outcomes to guide subsequent measurements even when quantum memory is reset after each pair.

Beyond estimating the purity of a single state, one may wish to identify the purest among several unknown states, for example, when comparing the outputs of different quantum state preparation procedures or devices~\citep{elben-2020-cross}.
\citet{yu2025pqsi} introduced purest quantum state identification under a fixed sample budget and explicitly left the fixed-confidence setting as an open problem.
We give a high-probability guarantee of correct identification within a gap-dependent number of copies under limited quantum memory.
The learner must adaptively allocate samples among competing states and decide when to stop, without prior knowledge of the purity gaps.

\subsection{Main Results}
\noindent\textbf{Purity estimation.}
Our first result closes the gap in the sample complexity of purity estimation with limited quantum memory.
Let \(d=2^n\) and \(r=2^k\), where \(r\) is the dimension of the quantum register retained between two copies.
We prove that, for every \(0\le k\le n\), the optimal worst-case expected sample complexity is \(\Theta(\Rrate{\eps})\), where
\begin{equation}\label{eq:rate}
\Rrate{\eps}=\max\ab\{\eps^{-2},\min\ab\{\sqrt d/\eps,d/(r\eps^2)\}\}.
\end{equation}
This result holds for sufficiently small \(\eps\) and constant success probability.
Our lower bound matches the existing upper bound of \citet{gong}, even when the learner can adapt its measurements based on all previous outcomes and stop at a random time.
The key to our proof is a new bound on the Fisher information contained in the entire adaptive measurement record (Lemma~\ref{lem:information}).
Table~\ref{tab:purity-comparison} compares our result with prior work.

\begin{table}[htbp]
\caption{Purity estimation complexity with and without quantum memory.
Adaptivity refers to dependence on earlier rounds' classical outcomes.}
\label{tab:purity-comparison}
\centering
\small
\setlength{\extrarowheight}{0pt}
\addtolength{\extrarowheight}{\aboverulesep}
\addtolength{\extrarowheight}{\belowrulesep}
\setlength{\aboverulesep}{0pt}
\setlength{\belowrulesep}{0pt}
\begin{threeparttable}
\begin{tabular}{llll}
\toprule
\textbf{Quantum memory}                                                                                                                           & \textbf{Reference}                                                                                                                     & \textbf{Adaptivity} & \textbf{Sample complexity}                                                                    \\
\midrule
\multirow{2}{*}{\begin{tabular}[c]{@{}l@{}}No memory\\\((r=1)\)\end{tabular}}                                                                       & \citet{all}\tnote{\(\dagger\)}                                                                                                          & Nonadaptive         & \(O\ab(\max\ab\{\eps^{-2},\frac{\sqrt d}{\eps}\})\)                                           \\
                                                                                                                                                  & \citet{gong}                                                                                                                           & Unrestricted        & \(\Omega\ab(\max\ab\{\eps^{-2},\frac{\sqrt d}{\sqrt{\eps}}\})\)                               \\
\cmidrule{1-4}
\multirow{2}{*}{\begin{tabular}[c]{@{}l@{}}With memory\\\((2\le r\le d)\)\end{tabular}}                                                             & \citet{gong}                                                                                                                           & Nonadaptive         & \(O\ab(\max\ab\{\eps^{-2},\min\ab\{\frac{\sqrt d}{\eps},\frac{d}{r\eps^2}\}\})\)              \\
                                                                                                                                                  & \citet{gong}                                                                                                                           & Unrestricted        & \(\Omega\ab(\max\ab\{\eps^{-2},\min\ab\{\frac{\sqrt d}{\sqrt{\eps}},\frac{d}{r\eps^2}\}\})\)  \\
\noalign{\kern-\cmidrulewidth}\cmidrule{1-4}
\rowcolor[rgb]{0.871,0.867,0.855} \begin{tabular}[c]{@{}>{\cellcolor[rgb]{0.871,0.867,0.855}}l@{}}All memory sizes\\\((1\le r \le d)\)\end{tabular} & \begin{tabular}[c]{@{}>{\cellcolor[rgb]{0.871,0.867,0.855}}l@{}}This work\\(Theorem~\ref{thm:complexity})\end{tabular} & Unrestricted        & \(\Theta\ab(\max\ab\{\eps^{-2},\min\ab\{\frac{\sqrt d}{\eps},\frac{d}{r\eps^2}\}\})\)    \\
\bottomrule
\end{tabular}
\begin{tablenotes}
\footnotesize
\item[\(\dagger\)]
\citet{all} also prove a matching lower bound for distributed inner product estimation.
We do not list it here, since it does not directly apply to purity estimation.
\end{tablenotes}
\end{threeparttable}
\end{table}

\noindent\textbf{Purest state identification.}
We further study the problem of identifying the purest among \(K\) unknown quantum states with probability at least \(1-\delta\).
Let \(i^\star\) denote the unique purest state and \(\Delta=\min_{i\ne i^\star}\{P(\vec\rho_{i^\star})-P(\vec\rho_i)\}\) be the minimum purity gap.
We develop a confidence elimination algorithm that adaptively allocates copies among the states.
Together with a matching lower bound, we establish that the minimax high-probability sample complexity, at constant confidence and sufficiently small \(\Delta\), is
\begin{equation}\label{eq:id-rate-intro}
\widetilde{\Theta}\ab(
K\max\ab\{
\Delta^{-2},
\min\ab\{\sqrt d/\Delta,d/(r\Delta^2)\}
\}
).
\end{equation}
Theorem~\ref{thm:id-upper} bounds the number of copies and guarantees correct identification on the same event of probability at least \(1-\delta\).
Theorem~\ref{thm:id-lower} gives a matching high-probability budget lower bound, up to logarithmic factors at constant confidence, for any sufficiently small collection of gaps.
The lower bound uses a single hard distribution to combine sourcewise sampling probabilities into a lower bound on the total copy budget.
Table~\ref{tab:identification-comparison} compares our results with prior work on purest quantum state identification.

\begin{table}[htbp]
\caption{High-probability copy budgets for purest state identification over instances with minimum purity gap at least \(\Delta\).
The minimax bounds match up to logarithmic factors at constant confidence and sufficiently small \(\Delta\).
Each round processes two copies of one selected source and resets quantum memory after the pair.}
\label{tab:identification-comparison}
\centering
\small
\setlength{\extrarowheight}{0pt}
\addtolength{\extrarowheight}{\aboverulesep}
\addtolength{\extrarowheight}{\belowrulesep}
\setlength{\aboverulesep}{0pt}
\setlength{\belowrulesep}{0pt}
\resizebox{\linewidth}{!}{%
\begin{threeparttable}
\begin{tabular}{llll}
\toprule
\textbf{Quantum memory}                                                          & \textbf{Reference}                                                                                                                                                       & \textbf{Copy budget}                                                                                                                                                              & \textbf{Assumptions}                                                                                                                                    \\
\midrule
\begin{tabular}[c]{@{}l@{}}Without memory\\\((r=1)\)\end{tabular}           & \citet{yu2025pqsi}\tnote{\(\dagger\)}                                                                                                                                            & \(\widetilde{O}\ab(K\max\ab\{\frac{1}{\Delta^2},\frac{d}{\Delta^{3/2}}\})\)                                                                                                   & Known gap lower bound                                                                                                                                      \\
\cmidrule{1-4}
\begin{tabular}[c]{@{}l@{}}Full memory\\\((r=d)\)\end{tabular}             & \citet{yu2025pqsi}\tnote{\(\dagger\)}                                                                                                                                            & \(\widetilde{O}\ab(\frac{K}{\Delta^2})\)                                                                                                                                      & Known gap lower bound                                                                                                                                      \\
\noalign{\kern-\cmidrulewidth}\cmidrule{1-4}
\rowcolor[rgb]{0.871,0.867,0.855} \begin{tabular}[c]{@{}>{\cellcolor[rgb]{0.871,0.867,0.855}}l@{}}All memory sizes\\\((1\le r\le d)\)\end{tabular} & \begin{tabular}[c]{@{}>{\cellcolor[rgb]{0.871,0.867,0.855}}l@{}}This work\\(Theorems~\ref{thm:id-upper} and~\ref{thm:id-lower})\end{tabular} & \(\widetilde{\Theta}\ab(K\max\ab\{\frac{1}{\Delta^2},\min\ab\{\frac{\sqrt d}{\Delta},\frac{d}{r\Delta^2}\}\})\) & No gap knowledge required \\
\bottomrule
\end{tabular}
\begin{tablenotes}
\footnotesize
\item[\(\dagger\)]
\citet{yu2025pqsi}'s upper bounds are converted from fixed-budget error guarantees using a known gap lower bound.
\end{tablenotes}
\end{threeparttable}
}
\end{table}

\subsection{Overview of Techniques}

\noindent\textbf{Purity estimation.}
The main difficulty is that resetting quantum memory does not erase the learner's classical knowledge.
Earlier outcomes can reveal information about the unknown state and guide later measurements, even though no quantum system survives between pairs.
The lower bound in \citet{gong} already permits this adaptation.
However, their comparison of the complete likelihood with a product over pairs incurs a correction of order \(J^2\eps/d\) after \(J\) pairs.
Keeping this correction small limits that part of the argument to the \(\sqrt{d/\eps}\) scale, rather than the desired \(\sqrt d/\eps\) scale.

We instead compare nearby mixture weights in the family
\(\vec\rho_{\theta,\vec\psi}=(1-\theta)\vec I/d+\theta\ket{\vec\psi}\bra{\vec\psi}\),
which also appears in \citet{yu2025pqsi}.
We choose two values of \(\theta\) so that the corresponding purities differ by order \(\eps\).
The direction \(\ket{\vec\psi}\) is drawn once from the Haar distribution and remains fixed throughout the experiment, creating dependence across measurement rounds.
As the learner adapts its measurements based on earlier outcomes, the posterior distribution of \(\ket{\vec\psi}\) generally differs from its initial Haar distribution.
Consequently, information bounds obtained by averaging a fixed measurement over the Haar distribution cannot be directly applied to subsequent measurements.

Our key technical step is a spherical score identity for the complete adaptive measurement record.
For each fixed record, the measurements selected along that record are fixed, allowing us to relate derivatives with respect to \(\theta\) to derivatives along the hidden direction through spherical integration by parts.
The resulting identity expresses the score of the marginal record distribution in terms of contributions within individual pairs and interactions across rounds.
By operating on the full likelihood, it accounts for the dependence induced by earlier observations without explicitly tracking the evolving posterior of \(\ket{\vec\psi}\).

We bound these contributions using local inequalities that hold uniformly over all histories and hidden directions.
Within each pair, compression of the first input into an \(r\)-dimensional register yields a rank constraint that bounds the corresponding second moment by \(O(dr)\).
Across rounds, a directional covariance bound controls interactions with the accumulated history without introducing an additional factor of \(d\).
Conditioned on the hidden direction, the contributions form a martingale difference sequence, allowing their second moments to be summed despite adaptive measurement choices.
Combining these bounds yields
\[
I(Q_\theta)
\le C\min\ab\{J,\frac{J^2+Jr}{d}\},
\]
where \(Q_\theta\) is the law of the complete record after averaging over the fixed hidden direction \(\ket{\vec\psi}\), and \(I(Q_\theta)\) is its Fisher information with respect to \(\theta\).
The terms \(J^2/d\) and \(Jr/d\) capture information accumulated across rounds and within pairs, respectively.

Integrating this information bound between the two nearby parameters shows that a constant distinguishing advantage requires both
\(J=\Omega(\eps^{-2})\) and \(J^2+Jr=\Omega(d/\eps^2)\).
These inequalities yield the optimal copy lower bound.
Thus, the sharp accuracy dependence follows from controlling the sensitivity of the full adaptive record between nearby interior parameters.
By declaring the alternative whenever a truncation limit is reached, we control the loss of distinguishing advantage using only the expected cost under the base hypothesis.
This testing reduction handles random stopping and yields a purity lower bound averaged under the fixed prior at \(\theta=1/2\).

\noindent\textbf{Purest state identification.}
Confidence elimination uses the established purity estimators to construct intervals and remove sources whose upper bounds fall below another source's lower bound.
A conditional union bound makes every requested estimate accurate on one event of probability at least \(1-\delta\).
On this event, the best source survives, each challenger is removed at its own gap scale, and the geometrically growing epoch budgets give the individual-gap sample bound.

For the lower bound, we construct a single base prior consisting of one reference state and \(K-1\) less pure challengers with independent hidden Haar directions.
Each alternative changes one challenger to make it the purest.
The same-source reset structure permits classical simulation of all other sources.
Capping queries to the changed source converts a low sampling probability into a test that contradicts the Fisher information bound.
A bounded weighted sum combines these sourcewise probability bounds into a constant-probability lower bound on total copies under the common prior.
A quantum relative entropy argument supplies the confidence-dependent term by bounding the information gained from each queried pair.

\subsection{Related Work}

\noindent\textbf{Coherent estimation of nonlinear functionals.}
Joint measurements on multiple copies allow nonlinear state properties to be estimated without reconstructing the density matrix, including purity and overlaps \citep{ekert2002direct} and more general polynomial functionals \citep{brun2004polynomial}.
\citet{liuwang2026powers} study trace powers under coherent state preparation oracles and unrestricted sample access, while \citet{chen2026simultaneous} estimate multiple nonlinear functionals simultaneously using collective measurements.
The purity comparisons considered by \citet{liuwang2026powers} take state preparation circuits as input.
Our identification problem instead concerns allocating samples among unknown states under a quantum memory constraint.

\noindent\textbf{Randomized measurements and classical shadows.}
\citet{elben2019randomized} recover purity and overlaps from correlations of randomized outcomes, using global or local random unitaries.
\citet{huang2020shadows} develop classical shadows for predicting observables and unbiased estimators of polynomial functionals from independent records.
\citet{elben2020mixed} use these records to estimate partially transposed moments and detect mixed-state entanglement.
\citet{du} optimize randomized measurements for a family of nonlinear properties.
\citet{li2026singlemeasurement} estimate higher moments and functionals \(\tr(\vec O\vec\rho^t)\), reducing the number of measurement settings to one in sufficiently large dimension or by adding fresh pure ancillas.
Their lower bound \(\Omega(d^{1-1/t})\), however, assumes repeated measurement in one fixed orthonormal basis at constant accuracy, rather than arbitrary adaptive POVMs.

\noindent\textbf{Quantum memory and purity testing.}
Quantum memory can substantially reduce the sample complexity of learning quantum states \citep{aharonov2022qualm,chenmemory}.
\citet{chen2024tradeoffs} establish the optimal memory tradeoff for purity testing, which distinguishes a pure state from the maximally mixed state.
In the two-copy reset model with \(r=2^k\) dimensional quantum memory, they obtain the tight sample complexity
\(\Theta(\min\{\sqrt d,d/r\})\).
This characterizes the memory tradeoff for a testing problem with constant purity separation, but does not determine the dependence on additive estimation accuracy.
\citet{akresh} use a relaxation to positive partial transpose measurements to give a simpler proof of the tight \(\Omega(\sqrt d)\) lower bound for purity testing with adaptive single-copy measurements.
\citet{gong} subsequently study purity estimation in the same memory model as \citet{chen2024tradeoffs}, establishing an upper bound for general accuracy and a lower bound that permits adaptive measurements but leaves a gap in its accuracy dependence.
We close the remaining gap for arbitrary classical adaptation and all permitted memory sizes.

\noindent\textbf{Accuracy and adaptive estimation of state moments.}
\citet{leewadhwa2026purity} characterize the optimal sample complexity of purity estimation under a multiplicative error guarantee, requiring \(|\widehat P-P(\vec\rho)|\le\alpha P(\vec\rho)\), using adaptive single-copy measurements.
Unlike our additive error guarantee, their permitted error depends on the unknown purity and can be as small as \(\alpha/d\) for nearly maximally mixed states.
\citet{liu2026moments} determines the optimal dimension dependence for adaptive estimation of state moments \(\tr(\vec\rho^t)\) at sufficiently small constant additive error.
For purity, their result gives \(\Theta(\sqrt d)\) copies, while higher moments exhibit different rates and advantages from adaptive measurements.
Our result instead determines the joint dependence on additive accuracy and retained quantum memory.

\noindent\textbf{Purest state identification and classical pure exploration.}
\citet{yu2025pqsi} introduce purest quantum state identification with a fixed sample budget, develop coherent and incoherent algorithms, and establish a lower bound for incoherent measurements using a randomly chosen POVM with two outcomes that remains fixed throughout the procedure.
We give high-probability sample complexity bounds across quantum memory sizes without prior gap knowledge.
\citet{ng2026spectral} studies quantum bandits with known linear reward observables and cumulative regret.
Our upper bound builds on confidence elimination \citep{evendar2006elimination}, while classical identification provides change of measure and allocation tools \citep{kaufmann2016complexity,garivier2016optimal}.
When SWAP measurements are available, their outcomes yield a Bernoulli identification problem.
Under restricted memory, our algorithm uses batch purity estimates.
Our lower bound combines sampling-probability requirements for every challenger under one common distribution of instances, producing an additive lower bound on the high-probability copy budget.

\section{Problem Setup and Measurement Model}\label{sec:model}

A quantum state on \(\CC^d\) is an operator \(\vec\rho\succeq0\) with \(\tr(\vec\rho)=1\).
A positive operator valued measure (POVM) has effects \(\vec M_y\succeq0\) summing to \(\vec I\), with outcome probabilities \(\tr(\vec M_y\vec\rho)\).
A quantum instrument is a collection of completely positive, trace-nonincreasing maps \(\{\mathcal I_s\}_s\) whose sum is trace preserving.
For discrete outcomes, \(\mathcal I_s(\vec\rho)=\sum_a\vec M_{s,a}\vec\rho\vec M_{s,a}^\dagger\) with \(\sum_{s,a}\vec M_{s,a}^\dagger\vec M_{s,a}=\vec I\), and outcome \(s\) has probability \(\tr(\mathcal I_s(\vec\rho))\).
All classical outcome spaces are standard Borel, and all policies are measurable.
Appendix~\ref{sec:regularity} constructs the reference densities for general outcomes.
For a random variable \(X\) in an experiment indexed by \(\xi\), write \(P_\xi^X:=\operatorname{Law}_\xi(X)\) for its distribution.

\subsection{Two-Copy Reset Quantum Memory}

We adopt the two-copy reset quantum memory model of \citet{gong}.
Let \(d=2^n\) be the dimension of the unknown quantum state and \(r=2^k\) the dimension of the retained quantum register, where \(n\ge1\) and \(0\le k\le n\).
In each round, the learner first applies a quantum instrument to a fresh copy of the state, records the classical outcome, and retains a quantum register with dimension \(\le r\).
It then measures the retained register jointly with a second fresh copy and discards all quantum systems before the next round.
Thus, quantum information can be retained between two copies within a round, but not across different rounds.

The learner may use unrestricted classical memory, computation, randomization, and adaptive measurement and stopping rules.
Its decisions may depend on the entire classical history, but not directly on the unknown state.
Additional ancillas are allowed during each operation, subject to the restriction on the retained quantum register.
Fresh ancillas measured with an individual input, as in \citet{li2026singlemeasurement}, implement a POVM on that input and do not increase the retained dimension \(r\).
When \(r=1\), no quantum information can be retained, and the model includes arbitrary adaptive single-copy measurements.

\subsection{Estimation and Identification Problems}
For purity estimation, an algorithm \(A\) receives copies of an unknown state \(\vec\rho\) and outputs an estimate \(\widehat P\) of its purity \(P(\vec\rho)\).
Let \(T_A\) denote the total number of copies consumed by \(A\).
We call \(A\) admissible if, for every state \(\vec\rho\), \(\Pr_{\vec\rho}(T_A<\infty)=1\) and \(\Pr_{\vec\rho}(|\widehat P-P(\vec\rho)|\le\eps)\ge3/4\).
The success probability \(3/4\) can be replaced by any fixed constant greater than \(1/2\) without changing the sample complexity beyond constant factors, through independent repetitions and median aggregation.

For purest state identification, consider \(K\ge2\) unknown states \(\nu=(\vec\rho_1,\ldots,\vec\rho_K)\), with purities \(\lambda_i:=P(\vec\rho_i)\).
We assume a unique purest state, indexed by \(i^\star=\argmax_{i \in [K]}\lambda_i\), and define purity gaps \(\Delta_i:=\lambda_{i^\star}-\lambda_i>0\) for \(i\ne i^\star\).
In each round \(t\), the learner selects a source \(I_t\in\{1,\ldots,K\}\) based on its classical history and processes two fresh copies of \(\vec\rho_{I_t}\) under the same reset memory constraint.
The learner may retain its entire classical history but must discard all quantum information after each pair and stop only at pair boundaries.
Let \(T_A\) be the total copy count and let \(\widehat i_A\) be the source index returned by \(A\).
We call \(A\) \(\delta\)-correct on a class of instances if \(\Pr_\nu(T_A<\infty,\widehat i_A=i^\star)\ge1-\delta\) for every instance \(\nu\) in that class.
A high-probability sample bound \(B(\nu,\delta)\) guarantees \(\Pr_\nu(T_A\le B(\nu,\delta),\widehat i_A=i^\star)\ge1-\delta\), so correctness and the copy bound hold on the same event.

\section{Purity Estimation}\label{sec:estimation}

In this section, we establish the optimal sample complexity of purity estimation in the two-copy reset memory model, allowing arbitrary classical adaptation and random stopping.

\subsection{Optimal Sample Complexity}\label{sec:estimation-complexity}

Theorem~\ref{thm:complexity} characterizes the optimal sample complexity for all permitted quantum memory sizes.

\begin{theorem}[Optimal sample complexity]\label{thm:complexity}
There exists a universal \(\eps_0>0\) such that, for \(d=2^n\), \(r=2^k\), integers \(n\ge1\) and \(0\le k\le n\), and \(0<\eps\le\eps_0\),
\[
\inf_{A\text{ admissible}}\sup_{\vec\rho}\E_{\vec\rho}T_A=\Theta\ab(\max\ab\{\eps^{-2},\min\ab\{\sqrt d/\eps,d/(r\eps^2)\}\}).
\]
\end{theorem}

The upper bound of Theorem~\ref{thm:complexity} follows from Theorem 8 and Claim 1 of \citet{gong}, with independent median amplification to achieve success probability \(3/4\).
Our contribution is the matching lower bound, stated in Theorem~\ref{thm:estimation-lower} and proved in Appendix~\ref{sec:estimation-lower}.

\noindent\textbf{Improvement over previous lower bounds.}
Our adaptive lower bound strengthens the \(\sqrt{d/\eps}\) term in \citet{gong} to \(\sqrt d/\eps\), matching the known upper bound for every permitted memory size.
The improvement follows from a new information bound for the complete adaptive measurement record, developed in Section~\ref{sec:lower} and Appendix~\ref{sec:estimation-lower}.

\noindent\textbf{Memory regimes.}
Table~\ref{tab:memory-regimes} summarizes the optimal sample complexity and the speedup relative to measurements without quantum memory (\(r=1\)).
When \(\eps\le d^{-1/2}\), even without quantum memory, the statistical limit \(\Theta(\eps^{-2})\) is achieved, so additional memory provides no asymptotic improvement.
For \(\eps>d^{-1/2}\), memory offers no improvement until \(r>\sqrt d/\eps\).
Beyond this threshold, the sample complexity decreases as \(\Theta(d/(r\eps^2))\), yielding a speedup of \(\Theta(r\eps/\sqrt d)\) over \(r=1\).
At \(r=d\), the complexity reaches \(\Theta(\eps^{-2})\).

\begin{table}[htbp]
\caption{Memory regimes in Theorem~\ref{thm:complexity}.}
\label{tab:memory-regimes}
\centering
\small
\begin{tabular}{llll}
\toprule
Accuracy & Register dimension & Sample complexity & Speedup over \(r=1\) \\
\midrule
\(\eps\le d^{-1/2}\)
    & \(1\le r\le d\)
    & \(\Theta(\eps^{-2})\)
    & \(\Theta(1)\) \\
\midrule
\multirow{2}{*}{\(\eps>d^{-1/2}\)}
    & \(1\le r\le\sqrt d/\eps\)
    & \(\Theta(\sqrt d/\eps)\)
    & \(\Theta(1)\) \\
\cmidrule(){2-4}
    & \(\sqrt d/\eps<r\le d\)
    & \(\Theta(d/(r\eps^2))\)
    & \(\Theta(r\eps/\sqrt d)\) \\
\bottomrule
\end{tabular}
\end{table}

\subsection{Matching Lower Bound under Adaptive Measurements}\label{sec:lower}

\subsubsection{Hard Instance and Information Bound}

Draw a Haar unit vector \({\vec{\psi}}\) once and consider
\begin{equation}\label{eq:states}
{\vec{\rho}}_{\theta,{\vec{\psi}}}=(1-\theta){\vec{I}}/d+\theta\ket{\vec{\psi}}\bra{\vec{\psi}},\qquad\theta\in[1/3,2/3].
\end{equation}
Write \({\vec{P}}_{\vec{\psi}}=\ket{\vec{\psi}}\bra{\vec{\psi}}\).
The purity of the constructed state is
\begin{equation}\label{eq:purity-family}
P_d(\theta)=\tr({\vec{\rho}}_{\theta,{\vec{\psi}}}^2)=\frac1d+\ab(1-\frac1d)\theta^2.
\end{equation}
Thus estimating purity distinguishes nearby values of \(\theta\), even though the learner does not know \({\vec{\psi}}\).

Let \(H\) be the complete classical record of a procedure using \(J\) rounds.
Include the learner's random seed in this record.
Let \(\sigma\) be Haar probability measure on the unit sphere and \(\Pr_{\theta,\vec\psi}\) the conditional record law.
Let \(A\) be any measurable set of possible complete records, so \(H\in A\) denotes the event that the realized record belongs to \(A\).
Let \(\mu:=\operatorname{Law}_{\vec I/d}(H)\) be the complete-record law of the same policy on maximally mixed inputs, including the seed, and define \(Q_\theta(A):=\int\Pr_{\theta,\vec\psi}(H\in A)\,\odif{\sigma}(\vec\psi)\) and \(q_\theta:=\odv{Q_\theta}{\mu}\).
Its Fisher information is
\begin{equation}\label{eq:fisher-definition}
I(Q_\theta)=\int\ab(\pdv{}{\theta}\log q_\theta(h))^2 q_\theta(h)\,\odif{\mu}(h).
\end{equation}
We justify the reference measure and the derivatives in Appendix~\ref{sec:regularity}.

\begin{restatable}[Information in the complete record]{lemma}{informationresult}\label{lem:information}
For the state family in Equation~\eqref{eq:states}, every procedure with \(J\) rounds and retained register dimension at most \(r\) satisfies
\[
I(Q_\theta)\le C\min\ab\{J,\frac{J^2+Jr}{d}\}
\qquad\text{for every }\theta\in[1/3,2/3],
\]
where \(C\) is universal.
The procedure may select all its operations using the complete classical history.
\end{restatable}

The record law integrates the shared Haar direction after forming the full adaptive likelihood, accounting for how past observations inform later measurements.
The \(J^2/d\) term allows correlations between observations from different rounds, while \(Jr/d\) bounds the additional information from comparisons using quantum memory.

\begin{restatable}[Sharpness of the information bound]{remark}{informationsharpness}
\label{rem:information-sharpness}
The information bound in Lemma~\ref{lem:information} is sharp up to universal constants, as shown by the matching construction in Appendix~\ref{app:information-sharpness}.
\end{restatable}

\subsubsection{A Spherical Identity for the Complete Record}

On the unit sphere \(S^{2d-1}\subset\RR^{2d}\), let \(\nabla\) and \(\mathcal L\) denote the gradient and Laplacian for the usual round metric.
Sphere derivatives act on \(\vec\psi\), holding \(\theta\), the classical history, and the selected measurement operations fixed.
For a smooth function \(f\) and a real tangent vector \(\vec v\), satisfying \(\operatorname{Re}\langle\vec\psi,\vec v\rangle=0\), define the directional derivative by
\[
\partial_{\vec v}f(\vec\psi):=\ab.\odv{f(\vec\gamma(t))}{t}|_{t=0},\qquad \vec\gamma(0)=\vec\psi,\quad \vec\gamma'(0)=\vec v,
\]
where \(\vec\gamma(t)\) is any smooth curve on the sphere with the stated initial point and velocity.
For matrix-valued functions, differentiation is applied entrywise.
Let \(\vec S\) exchange the two input systems and define
\begin{equation}\label{eq:contraction}
\vec K_{\vec\psi}=\ab(\vec P_{\vec\psi}\otimes(\vec I-\vec P_{\vec\psi})+(\vec I-\vec P_{\vec\psi})\otimes\vec P_{\vec\psi})\vec S.
\end{equation}
Coordinate differentiation gives
\begin{align}
\mathcal L\vec\rho_{\theta,\vec\psi}&=-4d\theta\,\pdv{\vec\rho_{\theta,\vec\psi}}{\theta},\label{eq:sphere-laplacian}\\
\sum_{\vec v}(\partial_{\vec v}\vec\rho_{\theta,\vec\psi})\otimes(\partial_{\vec v}\vec\rho_{\theta,\vec\psi})&=2\theta^2\vec K_{\vec\psi},\label{eq:sphere-contraction}
\end{align}
where the sum runs over an orthonormal real tangent basis at \(\vec\psi\).
Appendix~\ref{sec:sphere} supplies the coordinate calculation.
For a fixed history \(h\), let \(\nu_h\) be the next pair's outcome law on maximally mixed inputs and \(\vec F_y\) the POVM-effect density relative to \(\nu_h\).
Its conditional likelihood ratio is \(p_y=\tr(\vec F_y\vec\rho_{\theta,\vec\psi}^{\otimes2})\).
Define
\begin{equation}\label{eq:local-scores}
\vec g_y=\nabla\log p_y,
\qquad w_y=\frac{4\theta^2\tr(\vec F_y\vec K_{\vec\psi})}{p_y}.
\end{equation}
This density convention applies to both discrete and continuous outcomes.
For a realized record \(h=(\omega,y_1,\ldots,y_J)\), where \(\omega\) is the random seed, let \(\vec F_{i,h}\) be the effect density of outcome \(y_i\) under the round \(i\) measurement selected using \(\omega,y_1,\ldots,y_{i-1}\), and define
\[
p_i(h,\theta,\vec\psi):=\tr(\vec F_{i,h}\vec\rho_{\theta,\vec\psi}^{\otimes2}).
\]
This is the preceding \(p_y\) evaluated at that history and at \(y=y_i\), so \(p_i\) is the conditional likelihood ratio of the observed outcome relative to the maximally mixed reference kernel at the preceding history.
Similarly, \(\vec g_i,w_i\) are \(\vec g_y,w_y\) evaluated at the same history and outcome.

\begin{proposition}[Spherical score identity]\label{prop:score-identity}
For the complete record \(H\) of any \(J\)-round procedure, including its independent random seed,
\begin{equation}\label{eq:score-main}
\pdv{}{\theta}\log q_\theta(H)
=\frac{1}{4d\theta}\E_\theta[W\mid H],
\qquad W:=\sum_{i=1}^Jw_i+2\sum_{j<i}\vec g_i\cdot\vec g_j.
\end{equation}
The conditional expectation averages the single hidden direction under its posterior given the complete record.
\end{proposition}

\begin{proof}
Fix the record \(h\), including the seed.
All operations selected along this record are then fixed when differentiating \(\theta\) or \(\vec\psi\).
Resetting the register after every pair gives the conditional record likelihood ratio \(L_\theta(h,\vec\psi)=\prod_{i=1}^Jp_i(h,\theta,\vec\psi)\) relative to \(\mu\).
Equations~\eqref{eq:sphere-laplacian} and~\eqref{eq:sphere-contraction}, followed by the product rule, give
\begin{align}
\mathcal Lp_i&=-4d\theta\,\pdv{p_i}{\theta}+p_iw_i,\nonumber\\
\frac{\mathcal LL_\theta}{L_\theta}
&=\sum_i\frac{\mathcal Lp_i}{p_i}+2\sum_{j<i}\frac{\nabla p_i\cdot\nabla p_j}{p_ip_j}
=-4d\theta\,\pdv{}{\theta}\log L_\theta+W.\label{eq:product-main}
\end{align}
The sphere has no boundary, so \(\int\mathcal LL_\theta\,\odif{\sigma}=0\).
Multiplication by \(L_\theta\) and integration therefore yield
\[
4d\theta\,\pdv{q_\theta(h)}{\theta}
=\int L_\theta(h,\vec\psi)W(h,\theta,\vec\psi)\,\odif{\sigma}(\vec\psi).
\]
Divide by \(4d\theta q_\theta(h)\), using the posterior density \(L_\theta(h,\vec\psi)/q_\theta(h)\), to obtain Equation~\eqref{eq:score-main}.
The uniform density and derivative bounds in Appendix~\ref{sec:regularity} justify these steps for general measurable outcomes.
\end{proof}

The two terms in \(W\) require different controls.
The retained register bounds \(w_i\), while directional covariance controls the inner products between a new score and previous scores.

\subsubsection{Local Bounds and the Memory Dimension}

\begin{restatable}[Local measurement bounds]{lemma}{localresult}\label{lem:local}
For every fixed history and direction \(\vec\psi\), expectations over the next round satisfy
\begin{align*}
\E[\vec g_y]&=0,& \E[w_y]&=0,\\
\E[\vec g_y\vec g_y^{\mathsf T}]&\preceq C\vec I,&
\E\norm{\vec g_y}^2&\le Cd,&
\E[w_y^2]&\le Cdr.
\end{align*}
The constants are universal on \(\theta\in[1/3,2/3]\).
\end{restatable}

Both second-moment bounds follow from the same matrix inequality.
For a positive semidefinite \(\vec D\), an arbitrary operator \(\vec A\), and a POVM \(\{\vec B_y\}_y\),
\begin{equation}\label{eq:matrix-main}
\sum_y\frac{\ab(\tr[\vec B_y(\vec A\vec D+\vec D\vec A^\dagger)])^2}{\tr(\vec B_y\vec D)}\le4\tr(\vec A\vec D\vec A^\dagger).
\end{equation}
Here \(0/0:=0\).
The inequality follows from trace Cauchy and POVM completeness, as detailed in Lemma~\ref{lem:matrix}.
For effect densities, the same inequality holds with the sum replaced by an integral against the reference outcome measure.
The zero means in Lemma~\ref{lem:local} follow by differentiating POVM completeness and using \(\tr(\vec K_{\vec\psi})=0\).

Applied to the derivative of the uncompressed pair state along \(\vec v\in\vec\psi^\perp\), Equation~\eqref{eq:matrix-main} gives
\[
\E[(\vec v\cdot\vec g_y)^2]=\int\frac{(\partial_{\vec v}p_y)^2}{p_y}\,\odif{\nu_h}(y)\le8\theta\norm{\vec v}^2.
\]
The phase direction \(i\vec\psi\) has zero derivative, so this controls every real tangent direction and yields \(\E[\vec g_y\vec g_y^{\mathsf T}]\preceq C\vec I\).
Taking the trace gives \(\E\norm{\vec g_y}^2\le Cd\).
Thus each direction carries only constant information, even though there are order \(d\) directions.

For the memory term, refine the first instrument into Kraus maps \(\vec M_s:\CC^d\to\CC^r\) with \(\sum_s\vec M_s^\dagger\vec M_s=\vec I\).
This refinement can only increase the second moment, so it suffices to bound the refined measurement.
The orthogonal component retained in branch \(s\) is described by the positive operator
\[
\vec C_s:=\vec M_s(\vec I-\vec P_{\vec\psi})\vec M_s^\dagger,\qquad \rank(\vec C_s)\le r.
\]
Applying Equation~\eqref{eq:matrix-main} to the compressed branches gives
\begin{equation}\label{eq:rank-main}
\E[w_y^2]\le Cd\sum_s\rank(\vec C_s)\norm{\vec M_s\vec\psi}^2\le Cdr,
\qquad \sum_s\norm{\vec M_s\vec\psi}^2=1.
\end{equation}
The factor \(d\) comes from the eigenvalue \((1-\theta)/d\) of the input state on \(\vec\psi^\perp\), while compression limits the contributing rank to \(r\).
Instrument completeness removes any dependence on the number of branches.
Appendix~\ref{sec:local-details} supplies the operator constructions, the calculation on singular supports, and the treatment of zero likelihoods and continuous outcomes.

\subsubsection{Controlling Information Accumulation under Adaptation}

Condition on the hidden direction and let \(\mathcal F_i=\sigma(\omega,Y_1,\ldots,Y_i)\), where \(\omega\) is the seed and \(Y_i\) is the complete outcome of pair \(i\).
Set \(\vec V_{i-1}=\sum_{j<i}\vec g_j\).
Lemma~\ref{lem:local} gives conditional mean zero for \(\vec g_i,w_i\), so martingale orthogonality yields \(\E[\norm{\vec V_{i-1}}^2\mid\vec\psi]\le Cd(i-1)\).
The directional covariance bound gives the crucial conditional estimate
\begin{align*}
\E[(\vec g_i\cdot\vec V_{i-1})^2\mid\mathcal F_{i-1},\vec\psi]
=\vec V_{i-1}^{\mathsf T}\E[\vec g_i\vec g_i^{\mathsf T}\mid\mathcal F_{i-1},\vec\psi]\vec V_{i-1}
\le C\norm{\vec V_{i-1}}^2.
\end{align*}
Using only \(\E\norm{\vec g_i}^2\le Cd\) in a norm bound would lose an extra factor of \(d\).
The covariance bound directly controls the inner product with the history vector.
Since \(\xi_i=w_i+2\vec g_i\cdot\vec V_{i-1}\) also has conditional mean zero and \(W=\sum_i\xi_i\),
\begin{align}
\E[W^2\mid\vec\psi]
&=\sum_i\E[\xi_i^2\mid\vec\psi]
\le2\sum_i\E[w_i^2\mid\vec\psi]+8\sum_i\E[(\vec g_i\cdot\vec V_{i-1})^2\mid\vec\psi]\nonumber\\
&\le CdrJ+Cd\sum_i(i-1)\le Cd(Jr+J^2).\label{eq:martingale-main}
\end{align}
The bound allows arbitrary dependence between \(w_i\) and \(\vec g_i\) within a pair.
Proposition~\ref{prop:score-identity} and conditional Jensen now give
\[
I(Q_\theta)\le\frac{\E_\theta W^2}{16d^2\theta^2}\le C\frac{J^2+Jr}{d}.
\]
For the other branch, \(-3\vec\rho\preceq\pdv{\vec\rho}{\theta}\preceq3\vec\rho\) implies that the radial score \(s_i=\pdv{}{\theta}\log p_i\) has \(|s_i|\le6\) and conditional mean zero.
Hence \(\E_\theta[(\sum_i s_i)^2]\le36J\), and averaging out the direction only decreases this second moment.
This proves Lemma~\ref{lem:information}.
Appendix~\ref{sec:information-details} gives the full regularity argument and details of the conditional expectations.

\subsubsection{From Information to Sequential Testing and Estimation}

For probability measures \(Q,Q'\) on the same measurable space, with densities \(q,q'\) relative to a common dominating measure \(\mu\), define their total variation distance by
\[
\TV(Q,Q'):=\sup_{A\text{ measurable}}\ab|Q(A)-Q'(A)|=\frac12\int\ab|q(h)-q'(h)|\,\odif{\mu}(h).
\]

For the square root densities of a fixed-budget experiment, the fundamental theorem of calculus and the Cauchy inequality give
\begin{equation}\label{eq:testing-main}
\TV(Q_{\theta_0},Q_{\theta_1})\le\frac12\int_{\theta_0}^{\theta_1}\sqrt{I(Q_\theta)}\,\odif{\theta}.
\end{equation}
Let \(h:=\theta_1-\theta_0\), where \(1/3\le\theta_0<\theta_1\le2/3\), and let \(T_{\mathrm{test}}\) count copies from the tested source.
For sufficiently small \(h\), a sequential test in the reset model with one fixed Haar direction, almost sure termination and error at most \(1/16\) at each endpoint satisfies
\[
\E_{\theta_0}T_{\mathrm{test}}\ge c\Rrate{h}.
\]
The expectation includes the direction and the tester's randomness.

For finite base expectation, cap the number of queried pairs at a constant multiple of \(\E_{\theta_0}T_{\mathrm{test}}\).
Check normal stopping before the cap and declare the alternative when the cap is reached.
Markov's inequality bounds the increase in the base error, preserving a constant distinguishing advantage.
Padding to the deterministic pair budget and applying Lemma~\ref{lem:information} and Equation~\eqref{eq:testing-main} forces that budget to be at least order \(\Rrate{h}\).
Lemma~\ref{lem:one-sided-test} in Appendix~\ref{app:one-sided-test} gives the complete truncation argument.

For purity estimation, repetitions with independent seeds and fresh copies are conditionally independent given the fixed hidden direction.
Median aggregation over a constant number of repetitions reduces the error below the testing threshold.
Choose \(\theta_0=1/2\) and \(\theta_1=1/2+4\eps\), whose purities differ by more than \(2\eps\) for sufficiently small \(\eps\), and threshold the median estimate at their midpoint.
Lemma~\ref{lem:one-sided-test} then gives the lower bound under one fixed base prior:
\begin{equation}\label{eq:estimation-prior-main}
\E_{\vec\psi\sim\sigma}\E_{\vec\rho_{1/2,\vec\psi}}T_A\ge c\Rrate{\eps}.
\end{equation}
The worst-case lower bound follows immediately.
Appendix~\ref{sec:testing-details} gives the complete proof of Theorem~\ref{thm:estimation-lower}.

\section{Purest State Identification}
\label{sec:id-identification}

Adaptive allocation resolves each source at the accuracy set by its purity gap.
To include the best source in the upper bound, define its effective gap by
\[
\bar\Delta_i=\Delta_i\quad(i\ne i^\star),\qquad \bar\Delta_{i^\star}=\min_{i\ne i^\star}\Delta_i.
\]

\subsection{Upper Bound}
\label{sec:id-upper}

Algorithm~\ref{alg:id-elimination} repeatedly estimates the active sources and eliminates those whose confidence intervals lie below another source's.
The routine \(\textsc{Estimate}(i,a,\eta)\) estimates source \(i\)'s purity to accuracy \(a\) with failure probability at most \(\eta\), using the construction in Lemma~\ref{lem:estimate-primitive}.

\begin{algorithm}[H]
\caption{\textsc{Reset Purity Elimination}}
\label{alg:id-elimination}
\KwIn{Sources \(1,\ldots,K\), memory dimension \(r\), error budget \(\delta\).}
Initialize \(S=\{1,\ldots,K\}\)\;\label{line:id-init}
\For{\(j=1,2,\ldots\)}{
Set \(a_j=2^{-j}/4\)\;
\ForEach{\(i\in S\)}{
\(\widehat\lambda_i\gets\textsc{Estimate}(i,a_j,\delta/(2Kj^2))\)\;\label{line:id-estimate}
Set \(L_i=\widehat\lambda_i-a_j\) and \(U_i=\widehat\lambda_i+a_j\)\;
}
\(S\gets\{i\in S:U_i\ge\max_{h\in S}L_h\}\)\;\label{line:id-eliminate}
\If{\(\lvert S\rvert=1\)}{return its unique source\;\label{line:id-return}}
}
\end{algorithm}

\begin{restatable}[High-probability identification upper bound]{theorem}{identificationupper}
\label{thm:id-upper}
There is a universal constant \(C>0\) such that, for every \(0<\delta<1/2\) and every instance \(\nu\) with a unique purest source, Algorithm~\ref{alg:id-elimination} uses no prior gap knowledge and satisfies
\begin{equation}
\label{eq:id-upper}
\Pr_\nu\ab(\widehat i=i^\star,\ T\le C\sum_{i=1}^K\Rrate{\bar\Delta_i}\ab[\log\frac K\delta+\log\log\frac e{\bar\Delta_i}\ab])\ge1-\delta.
\end{equation}
The algorithm uses one \(k\)-qubit register in the reset model.
\end{restatable}

The per-call confidence levels in Line~\ref{line:id-estimate} make all requested estimates accurate simultaneously with probability at least \(1-\delta\).
On this event, elimination retains the best source and removes each challenger once the confidence radius reaches its gap scale (Line~\ref{line:id-eliminate}).
The algorithm returns the surviving source (Line~\ref{line:id-return}), and geometric summation of the epoch budgets gives Equation~\eqref{eq:id-upper}.
Appendix~\ref{app:id-upper} gives the complete proof.
At \(r=d\), SWAP outcomes are Bernoulli with means \((1+\lambda_i)/2\), reducing identification to a classical best-arm problem with dimension-independent copy complexity.

\subsection{Lower Bound}

We now establish a matching lower bound for purest state identification.
For \(0<\Delta\le1-1/d\), let \(\mathcal V_\Delta=\{\nu:\min_{i\ne i^\star}\Delta_i\ge\Delta\}\).
\begin{restatable}[High-probability identification lower bound]{theorem}{identificationlower}
\label{thm:id-lower}
There exist universal constants \(c,\Delta_0,\delta_0>0\) such that the following holds for \(d=2^n\), \(r=2^k\), integers \(n\ge1\), \(0\le k\le n\), \(K\ge2\), gaps \(0<\Delta_j\le\Delta_0\) for \(2\le j\le K\), and \(0<\delta\le\delta_0\).
Every algorithm that is \(\delta\)-correct on \(\mathcal V_\Delta\), where \(\Delta=\min_{2\le j\le K}\Delta_j\), has an instance \(\nu\) with \(i^\star=1\) and \(\lambda_1-\lambda_j=\Delta_j\) for all \(j\ge2\) such that
\begin{equation}
\label{eq:id-lower}
\Pr_\nu\ab(
T<\infty,\ \widehat i=i^\star,\
T\ge c\sum_{j=2}^K
\max\ab\{
\frac{\log(1/\delta)}{\Delta_j^2},
\Rrate{\Delta_j}
\}
)\ge\frac14.
\end{equation}
\end{restatable}

The proof constructs a common prior under which source \(1\) is optimal and the remaining sources have independent hidden Haar directions and prescribed purity gaps.
For each suboptimal source, an alternative changes only that source to make it optimal.
By limiting the number of copies requested from the changed source, we reduce identification to a testing problem.
The Fisher information bound from Section~3 and a quantum relative entropy argument establish the necessary sample costs under limited quantum memory and the desired confidence level, respectively.
Since these bounds hold under the same prior, they can be combined to obtain the lower bound on the total number of copies.
The complete proof is given in Appendix~\ref{app:id-lower}.

At constant confidence, Theorem~\ref{thm:id-lower} matches the upper bound in Theorem~5 up to logarithmic factors for any sufficiently small purity gaps.
The contribution of the optimal source to the upper bound is absorbed into those of the suboptimal sources, since
\(\Rrate{\bar\Delta_{i^\star}}
=\max_{j\ne i^\star}\Rrate{\Delta_j}
\le\sum_{j\ne i^\star}\Rrate{\Delta_j}\).
Moreover, since \(\delta\le\delta_0<1/4\), any valid high-probability sample budget must be at least the threshold in Equation~\eqref{eq:id-lower} on the constructed instance.
In particular, when all suboptimal sources have the same gap \(\Delta\), the minimax high-probability sample complexity over \(\mathcal V_\Delta\) is at least \(c(K-1)\max\ab\{\Delta^{-2}\log(1/\delta), \Rrate{\Delta}\}\), even when the algorithm knows \(\Delta\) in advance.

\section{Discussion}
\label{sec:discussion}
Our results focus on the two-copy reset model, where quantum information can be retained within a pair but not across measurement rounds.
An important open question is how persistent quantum memory affects the optimal sample complexity.
For purest state identification, we require each round to process two copies of the selected state.
Allowing joint measurements on copies from different states may lead to more efficient identification strategies.
The high-probability identification bounds characterize every sufficiently small gap profile up to logarithmic factors at constant confidence.
Their explicit confidence dependence separates scalar statistical costs from the costs of extracting purity information with limited memory.

\section{Conclusion}
We established the optimal sample complexity of purity estimation with limited quantum memory, closing the gap between the known upper and lower bounds under arbitrary classical adaptation.
Our analysis combines a spherical score identity, a memory-dependent rank bound, and a martingale argument to control information accumulated across measurement rounds.
Collision and block measurements attain the resulting information bound up to universal constants.
For identification, a single confidence-elimination procedure guarantees correct selection within a high-probability individual-gap sample budget without prior gap knowledge.
Matching probability lower bounds establish this gap dependence up to logarithmic factors for every sufficiently small gap profile at constant confidence.

\section*{AI Use Statement}
The authors used ChatGPT and Codex to assist in searching for and comparing related work, developing, verifying, and refining parts of the proofs, and improving the clarity of the manuscript.
The authors independently reviewed and verified all AI-assisted content incorporated into the manuscript and take full responsibility for its accuracy and correctness.

\bibliography{references}

\clearpage
\appendix
\section{Purity Estimation: Lower Bound}\label{sec:estimation-lower}

\subsection{Proof of Lemma~\ref{lem:local}}\label{sec:local-details}\label{sec:matrix}\label{sec:sphere}

\localresult*

This appendix proves the identities and local estimates used in Section~\ref{sec:lower}.
The dimension \(d\ge2\) is fixed, and \(\theta\in[1/3,2/3]\) throughout.
All constants in the final bounds are independent of \(d,r\), the history, and the chosen instruments.

\begin{lemma}\label{lem:matrix}
Let \({\vec{D}}\) be a positive semidefinite operator, let \({\vec{A}}\) be any operator on the same finite dimensional space, and let \(\{{\vec{B}}_y\}\) be a POVM.
Then
\begin{equation}\label{eq:matrix}
\sum_y\frac{\ab(\tr[{\vec{B}}_y({\vec{A}}{\vec{D}}+{\vec{D}}{\vec{A}}^\dagger)])^2}{\tr({\vec{B}}_y{\vec{D}})}\le4\tr({\vec{A}}{\vec{D}}{\vec{A}}^\dagger).
\end{equation}
Use the convention \(0/0:=0\).
The same conclusion holds with an integral for a POVM with continuous outcomes.
\end{lemma}

\begin{proof}
Define \({\vec{X}}={\vec{B}}_y^{1/2}{\vec{D}}^{1/2}\) and \({\vec{Y}}={\vec{B}}_y^{1/2}{\vec{A}}{\vec{D}}^{1/2}\).
The Cauchy inequality for the trace inner product gives
\begin{align*}
\ab|\tr({\vec{B}}_y{\vec{A}}{\vec{D}})|^2
&=\ab|\tr({\vec{X}}^\dagger {\vec{Y}})|^2\\
&\le\tr({\vec{X}}^\dagger {\vec{X}})\tr({\vec{Y}}^\dagger {\vec{Y}})
=\tr({\vec{B}}_y{\vec{D}})\tr({\vec{B}}_y{\vec{A}}{\vec{D}}{\vec{A}}^\dagger).
\end{align*}
Since \({\vec{B}}_y\) and \({\vec{D}}\) are Hermitian, \(\tr({\vec{B}}_y{\vec{D}}{\vec{A}}^\dagger)\) is the complex conjugate of \(\tr({\vec{B}}_y{\vec{A}}{\vec{D}})\).
It follows that
\[
\ab(\tr[{\vec{B}}_y({\vec{A}}{\vec{D}}+{\vec{D}}{\vec{A}}^\dagger)])^2
=4\ab(\operatorname{Re}\tr({\vec{B}}_y{\vec{A}}{\vec{D}}))^2
\le4\tr({\vec{B}}_y{\vec{D}})\tr({\vec{B}}_y{\vec{A}}{\vec{D}}{\vec{A}}^\dagger).
\]
If \(\tr({\vec{B}}_y{\vec{D}})=0\), the preceding bound makes the numerator zero as well.
For a positive denominator, divide by \(\tr({\vec{B}}_y{\vec{D}})\), then sum over \(y\).
The identity \(\sum_y{\vec{B}}_y={\vec{I}}\) gives Equation~\eqref{eq:matrix}.
The proof uses only positivity and completeness and therefore applies to effect densities followed by integration.
In particular, it does not require \({\vec{D}}\) to be invertible.
\end{proof}

Regard a complex vector \({\vec{\psi}}\) as a vector of \(2d\) real coordinates.
The real inner product is \(\langle\vec u,\vec v\rangle_{\RR}:=\operatorname{Re}\langle\vec u,\vec v\rangle\), so its unit sphere is \(S^{2d-1}\).
Let \(\sigma\) denote the normalized uniform measure on this sphere.
The operator \({\vec{P}}_{\vec{\psi}}=\ket{\vec{\psi}}\bra{\vec{\psi}}\) has entries \({\vec{\psi}}_a\overline{\vec{\psi}}_b\).
We compute their sphere Laplacian directly.

Let \(f_{ab}(\vec z):=z_a\overline z_b\) on \(\RR^{2d}\).
Writing each \(z_a\) in real and imaginary coordinates gives \(\Delta_{\RR^{2d}}f_{ab}=4\delta_{ab}\) and \(f_{ab}(t\vec z)=t^2f_{ab}(\vec z)\).
In polar coordinates with radius \(t\),
\[
\Delta_{\RR^{2d}}f=\pdv[2]{f}{t}+\frac{2d-1}{t}\pdv{f}{t}+\frac1{t^2}\mathcal L f.
\]
At \(t=1\), its radial derivatives are \(\pdv{f}{t}=2f\) and \(\pdv[2]{f}{t}=2f\).
Thus \(\mathcal L f_{ab}=4\delta_{ab}-4df_{ab}\), or
\begin{equation}\label{eq:projector-laplacian}
\mathcal L{\vec{P}}_{\vec{\psi}}=-4d\ab({\vec{P}}_{\vec{\psi}}-\frac {\vec{I}}d).
\end{equation}
Since \(\pdv{{\vec{\rho}}_{\theta,{\vec{\psi}}}}{\theta}={\vec{P}}_{\vec{\psi}}-{\vec{I}}/d\), multiplying Equation~\eqref{eq:projector-laplacian} by \(\theta\) proves Equation~\eqref{eq:sphere-laplacian}.

To derive the gradient identity, choose an orthonormal complex basis \({\vec{u}}_1,\ldots,{\vec{u}}_{d-1}\) of \({\vec{\psi}}^\perp\).
The vectors \({\vec{u}}_a,i{\vec{u}}_a\), together with \(i{\vec{\psi}}\), form an orthonormal real tangent basis.
For a tangent vector \({\vec{v}}\), differentiation along a smooth curve with initial velocity \({\vec{v}}\) gives
\[
\partial_{\vec{v}}{\vec{P}}_{\vec{\psi}}=\ket {\vec{v}}\bra{\vec{\psi}}+\ket{\vec{\psi}}\bra {\vec{v}}.
\]
Put \({\vec{A}}_a=\ket{{\vec{u}}_a}\bra{\vec{\psi}}\).
The directional derivatives are \(\partial_{\vec u_a}\vec P_{\vec\psi}=\vec A_a+\vec A_a^\dagger\) and \(\partial_{i\vec u_a}\vec P_{\vec\psi}=i(\vec A_a-\vec A_a^\dagger)\).
Expanding their tensor squares gives
\begin{align*}
&({\vec{A}}_a+{\vec{A}}_a^\dagger)\otimes({\vec{A}}_a+{\vec{A}}_a^\dagger)
-({\vec{A}}_a-{\vec{A}}_a^\dagger)\otimes({\vec{A}}_a-{\vec{A}}_a^\dagger)\\
&\hspace{1em}=2\ab({\vec{A}}_a\otimes {\vec{A}}_a^\dagger+{\vec{A}}_a^\dagger\otimes {\vec{A}}_a).
\end{align*}
The derivative in direction \(i{\vec{\psi}}\) is zero, because this direction only changes the global phase.
Since \(\sum_a\ket{\vec u_a}\bra{\vec u_a}=\vec I-\vec P_{\vec\psi}\) and \(\vec S\) exchanges tensor factors,
\[
\sum_a(\vec A_a\otimes\vec A_a^\dagger+\vec A_a^\dagger\otimes\vec A_a)=\ab(\vec P_{\vec\psi}\otimes(\vec I-\vec P_{\vec\psi})+(\vec I-\vec P_{\vec\psi})\otimes\vec P_{\vec\psi})\vec S=\vec K_{\vec\psi}.
\]
Multiplying the summed tensor derivatives by \(\theta^2\) therefore proves
\[
\sum_{\vec{v}}(\partial_{\vec{v}}{\vec{\rho}}_{\theta,{\vec{\psi}}})\otimes(\partial_{\vec{v}}{\vec{\rho}}_{\theta,{\vec{\psi}}})=2\theta^2{\vec{K}}_{\vec{\psi}}.
\]
For later use, \(\int\mathcal Lf\,\odif{\sigma}=0\) for every \(f\in C^\infty(S^{2d-1})\).
This follows by writing the Laplacian as the divergence of the sphere gradient and integrating on the compact sphere, which has no boundary.

\begin{proof}
Fix the history, \(\theta\), and \({\vec{\psi}}\), and abbreviate \({\vec{\rho}}={\vec{\rho}}_{\theta,{\vec{\psi}}}\).
The expectation in this proof is solely over the next round's outcomes.
We first use discrete outcomes for readability.
Outcomes whose effect is zero can be omitted.
For a discrete outcome \(y\), write \(\vec E_y:=\nu_h(\{y\})\vec F_y\) for its ordinary POVM effect and \(\pi_y:=\nu_h(\{y\})p_y=\tr(\vec E_y\vec\rho^{\otimes2})\) for its outcome probability.
The reference mass \(\nu_h(\{y\})\) is independent of \(\theta,\vec\psi\), so \(\vec g_y=\nabla\log\pi_y\) and \(w_y=4\theta^2\tr(\vec E_y\vec K_{\vec\psi})/\pi_y\).
Appendix~\ref{sec:regularity} supplies the same argument for effect densities.
Every remaining probability is positive because \({\vec{\rho}}\) is positive definite.

The mean of \(\vec g_y\) is
\[
\sum_y\pi_y\nabla\log\pi_y=\sum_y\nabla\pi_y=\nabla\sum_y\pi_y=0.
\]
Also, \(\tr({\vec{K}}_{\vec{\psi}})=0\), as is immediate from Equation~\eqref{eq:contraction} and \(\langle{\vec{\psi}},{\vec{u}}_a\rangle=0\).
Therefore
\[
\sum_y\pi_yw_y=4\theta^2\sum_y\tr({\vec{E}}_y{\vec{K}}_{\vec{\psi}})=4\theta^2\tr({\vec{K}}_{\vec{\psi}})=0.
\]

The eigenvalues of \({\vec{\rho}}\) in the direction of \({\vec{\psi}}\) and on its orthogonal complement are
\[
\lambda_0=\theta+\frac{1-\theta}{d},
\qquad \lambda_e=\frac{1-\theta}{d}.
\]
For a horizontal tangent vector \({\vec{v}}\in{\vec{\psi}}^\perp\), define \({\vec{a}}=(\theta/\lambda_0)\ket {\vec{v}}\bra{\vec{\psi}}\).
Using \({\vec{\rho}}\ket{\vec{\psi}}=\lambda_0\ket{\vec{\psi}}\), we obtain
\begin{align*}
{\vec{a}}{\vec{\rho}}+{\vec{\rho}} {\vec{a}}^\dagger
&=\theta\ab(\ket {\vec{v}}\bra{\vec{\psi}}+\ket{\vec{\psi}}\bra {\vec{v}})=\partial_{\vec{v}}{\vec{\rho}},\\
\tr({\vec{a}}{\vec{\rho}})&=\theta\langle{\vec{\psi}},{\vec{v}}\rangle=0,\\
\tr({\vec{a}}{\vec{\rho}} {\vec{a}}^\dagger)&=\frac{\theta^2}{\lambda_0}\norm {\vec{v}}^2\le\theta\norm {\vec{v}}^2.
\end{align*}
Let \({\vec{D}}={\vec{\rho}}\otimes{\vec{\rho}}\) and \({\vec{A}}={\vec{a}}\otimes {\vec{I}}+{\vec{I}}\otimes {\vec{a}}\).
The derivative of the two inputs is \(\partial_{\vec{v}}{\vec{D}}={\vec{A}}{\vec{D}}+{\vec{D}}{\vec{A}}^\dagger\).
The two cross terms in \(\tr({\vec{A}}{\vec{D}}{\vec{A}}^\dagger)\) each contain the factor \(\tr({\vec{a}}{\vec{\rho}})\) or its complex conjugate, and hence vanish.
The remaining terms give
\[
\tr({\vec{A}}{\vec{D}}{\vec{A}}^\dagger)=2\tr({\vec{a}}{\vec{\rho}} {\vec{a}}^\dagger)\tr({\vec{\rho}})\le2\theta\norm {\vec{v}}^2.
\]
Apply Lemma~\ref{lem:matrix} to the effects \({\vec{E}}_y\).
Since \(\partial_{\vec{v}}\pi_y=\pi_y({\vec{v}}\cdot\vec g_y)\),
\begin{equation}\label{eq:directional-bound}
\E[({\vec{v}}\cdot\vec g_y)^2]=\sum_y\frac{(\partial_{\vec{v}}\pi_y)^2}{\pi_y}\le8\theta\norm {\vec{v}}^2.
\end{equation}
In the remaining phase direction, \(\partial_{i\vec\psi}\vec P_{\vec\psi}=0\) and \((i\vec\psi)\cdot\vec g_y=0\).
For an arbitrary real tangent vector, its horizontal component has no greater norm, so Equation~\eqref{eq:directional-bound} proves the covariance inequality in Lemma~\ref{lem:local}.
Taking its trace in the \(2d-2\) horizontal directions yields \(\E\norm{\vec g_y}^2\le Cd\).

An instrument with retained output dimension at most \(r\) has a Kraus representation with maps from \(\CC^d\) to \(\CC^r\).
Pad a smaller output space to dimension \(r\), if necessary.
Refine the classical outcome to record an individual Kraus map, indexed by \(s\), so that
\[
{\vec{M}}_s:\CC^d\longrightarrow\CC^r,
\qquad \sum_s{\vec{M}}_s^\dagger {\vec{M}}_s={\vec{I}}.
\]
The later operations ignore any added refinement label and execute the original policy.
For each \(s\), write \(\{{\vec{B}}_{s,y}\}_y\) for the second POVM on \(\CC^r\otimes\CC^d\).
The unnormalized state it measures is \({\vec{\sigma}}_s=({\vec{M}}_s{\vec{\rho}} {\vec{M}}_s^\dagger)\otimes{\vec{\rho}}\).
The refined probability is \(p_{s,y}:=\tr(\vec B_{s,y}\vec\sigma_s)\).

The refinement is valid for the required upper bound on a quadratic likelihood expression.
Indeed, if probabilities \(p_a\) and real numerators \(t_a\) are merged into one outcome, scalar Cauchy gives
\begin{equation}\label{eq:merge}
\frac{(\sum_a t_a)^2}{\sum_a p_a}\le\sum_a\frac{t_a^2}{p_a}.
\end{equation}
Zero probability terms have zero numerators and are omitted.
The local numerator defining \(w_y\) is linear in the effect, so it merges by summation.
It suffices to bound the refined expression.

Define
\[
{\vec{x}}_s={\vec{M}}_s{\vec{\psi}},
\qquad {\vec{C}}_s={\vec{M}}_s({\vec{I}}-{\vec{P}}_{\vec{\psi}}){\vec{M}}_s^\dagger,
\qquad {\vec{D}}_s={\vec{C}}_s\otimes {\vec{P}}_{\vec{\psi}}.
\]
Expanding \({\vec{\rho}}=\lambda_0{\vec{P}}_{\vec{\psi}}+\lambda_e({\vec{I}}-{\vec{P}}_{\vec{\psi}})\) gives
\begin{align*}
{\vec{\sigma}}_s
&=\ab(\lambda_0\ket{{\vec{x}}_s}\bra{{\vec{x}}_s}+\lambda_e{\vec{C}}_s)
\otimes\ab(\lambda_0{\vec{P}}_{\vec{\psi}}+\lambda_e({\vec{I}}-{\vec{P}}_{\vec{\psi}}))\\
&\succeq\lambda_0\lambda_e{\vec{D}}_s.
\end{align*}
Every term in the tensor expansion is positive, which justifies the last inequality.
Set
\[
{\vec{T}}_s=\sum_{a=1}^{d-1}\ket{{\vec{x}}_s,{\vec{u}}_a}\bra{{\vec{M}}_s{\vec{u}}_a,{\vec{\psi}}}.
\]
Compressing the first factor in Equation~\eqref{eq:contraction} gives
\[
(\vec M_s\otimes\vec I)\vec K_{\vec\psi}(\vec M_s^\dagger\otimes\vec I)=\vec T_s+\vec T_s^\dagger.
\]
The product of these maps is
\begin{align}
{\vec{T}}_s^\dagger {\vec{T}}_s
&=\sum_{a,b}\ket{{\vec{M}}_s{\vec{u}}_a,{\vec{\psi}}}\langle {\vec{x}}_s,{\vec{u}}_a\mid {\vec{x}}_s,{\vec{u}}_b\rangle\bra{{\vec{M}}_s{\vec{u}}_b,{\vec{\psi}}}\nonumber\\
&=\norm{{\vec{x}}_s}^2\sum_a\ket{{\vec{M}}_s{\vec{u}}_a,{\vec{\psi}}}\bra{{\vec{M}}_s{\vec{u}}_a,{\vec{\psi}}}
=\norm{{\vec{x}}_s}^2{\vec{D}}_s.\label{eq:T-identity}
\end{align}
This calculation is valid even when \({\vec{x}}_s=0\) or \({\vec{D}}_s\) has a kernel.

Let \(\vec D_s^+\) be the Moore Penrose inverse and \(\vec\Pi_s:=\vec D_s^+\vec D_s\) its support projector.
Equation~\eqref{eq:T-identity} implies \({\vec{T}}_s={\vec{T}}_s{\vec{\Pi}}_s\), since \({\vec{D}}_s{\vec{v}}=0\) gives \(\norm{{\vec{T}}_s{\vec{v}}}^2=0\).
For \({\vec{A}}_s={\vec{T}}_s{\vec{D}}_s^+\), it follows that
\begin{align}
{\vec{A}}_s{\vec{D}}_s&={\vec{T}}_s{\vec{D}}_s^+{\vec{D}}_s={\vec{T}}_s,\nonumber\\
\tr({\vec{A}}_s{\vec{D}}_s{\vec{A}}_s^\dagger)
&=\tr({\vec{T}}_s{\vec{D}}_s^+{\vec{T}}_s^\dagger)
=\tr({\vec{D}}_s^+{\vec{T}}_s^\dagger {\vec{T}}_s)\nonumber\\
&=\norm{{\vec{x}}_s}^2\tr({\vec{D}}_s^+{\vec{D}}_s)
=\norm{{\vec{x}}_s}^2\rank({\vec{D}}_s)
\le r\norm{{\vec{x}}_s}^2.\label{eq:rank-charge}
\end{align}
The rank inequality follows from \(\rank(\vec D_s)=\rank(\vec C_s)\rank(\vec P_{\vec\psi})=\rank(\vec C_s)\le r\).

Apply Lemma~\ref{lem:matrix} with \({\vec{D}}_s,{\vec{A}}_s,{\vec{B}}_{s,y}\), then use \({\vec{\sigma}}_s\succeq\lambda_0\lambda_e{\vec{D}}_s\).
We obtain
\begin{align}
\sum_y\frac{\ab(\tr[{\vec{B}}_{s,y}({\vec{T}}_s+{\vec{T}}_s^\dagger)])^2}{\tr({\vec{B}}_{s,y}{\vec{\sigma}}_s)}
&\le\frac1{\lambda_0\lambda_e}\sum_y\frac{\ab(\tr[{\vec{B}}_{s,y}({\vec{A}}_s{\vec{D}}_s+{\vec{D}}_s{\vec{A}}_s^\dagger)])^2}{\tr({\vec{B}}_{s,y}{\vec{D}}_s)}\nonumber\\
&\le\frac{4r\norm{{\vec{x}}_s}^2}{\lambda_0\lambda_e}.\label{eq:compression-ratio}
\end{align}
If \(\tr({\vec{B}}_{s,y}{\vec{D}}_s)=0\), Lemma~\ref{lem:matrix} makes its numerator zero, so the comparison includes that case.
The refined value of \(w\) has numerator \(4\theta^2\tr[{\vec{B}}_{s,y}({\vec{T}}_s+{\vec{T}}_s^\dagger)]\).
Multiplying Equation~\eqref{eq:compression-ratio} by \(16\theta^4\) and summing over \(s\) therefore gives
\[
\E[w_y^2]\le\frac{64\theta^4r}{\lambda_0\lambda_e}\sum_s\norm{{\vec{M}}_s{\vec{\psi}}}^2.
\]
The sum equals \(\langle{\vec{\psi}},(\sum_s{\vec{M}}_s^\dagger {\vec{M}}_s){\vec{\psi}}\rangle=1\).
On the specified parameter interval, \(\lambda_0\ge1/3\) and \(\lambda_e\ge1/(3d)\), so \(\lambda_0\lambda_e\ge c/d\).
This proves \(\E[w_y^2]\le Cdr\).
Combining it with the means and directional estimates completes the lemma.
\end{proof}

\subsection{Proof of Lemma~\ref{lem:information}}\label{sec:information-details}\label{sec:regularity}

\informationresult*

\begin{proof}

Fix a realization \(\omega\) of the learner's random seed and a complete record \(h\) of its \(J\) rounds.
The seed distribution is independent of the state.
On this fixed record, the learner has selected definite operations at each round.
Their effects do not vary when we differentiate \(\theta\) or \({\vec{\psi}}\).
Use the conditional outcome kernels \(\nu_h\) of the same policy on maximally mixed inputs as reference measures, and let \(\mu\) be their complete record law, including the seed.
Let \({\vec{F}}_{i,h}\) be the effect density of the realized outcome in round \(i\) relative to its reference kernel, and put
\[
p_i(h,\theta,{\vec{\psi}})=\tr({\vec{F}}_{i,h}{\vec{\rho}}_{\theta,{\vec{\psi}}}^{\otimes2}).
\]
For discrete and general outcomes alike, \(p_i\) is the conditional likelihood ratio relative to the maximally mixed reference kernel.
For a nonzero discrete outcome \(y\) with ordinary effect \(\vec E_y\), this means \(\vec F_y=\vec E_y/[\tr(\vec E_y)/d^2]\) and \(p_y=\tr(\vec E_y\vec\rho^{\otimes2})/[\tr(\vec E_y)/d^2]\).
The reference kernels and densities for general outcomes are constructed at the end of this proof.
We suppress \(h,\theta,{\vec{\psi}}\) when there is no ambiguity.

Conditional on the one fixed direction \({\vec{\psi}}\), the inputs at every round are fresh independent copies of \({\vec{\rho}}_{\theta,{\vec{\psi}}}\).
The only information surviving from earlier rounds is the classical history.
Consequently, the conditional record likelihood ratio relative to \(\mu\) is
\begin{equation}\label{eq:record-product}
L_\theta(h,{\vec{\psi}})=\prod_{i=1}^Jp_i(h,\theta,{\vec{\psi}}).
\end{equation}
This is a product of conditional likelihood ratios, whose factors may depend on the entire preceding history.

The product rule for the sphere Laplacian and Equations~\eqref{eq:sphere-laplacian} and~\eqref{eq:sphere-contraction} give
\begin{align*}
\mathcal Lp_i
&=\tr\ab[{\vec{F}}_{i,h}\ab((\mathcal L{\vec{\rho}})\otimes{\vec{\rho}}+{\vec{\rho}}\otimes(\mathcal L{\vec{\rho}})+2\sum_v\partial_v{\vec{\rho}}\otimes\partial_v{\vec{\rho}})]\\
&=-4d\theta\tr\ab[{\vec{F}}_{i,h}\ab(\pdv{{\vec{\rho}}}{\theta}\otimes{\vec{\rho}}+{\vec{\rho}}\otimes\pdv{{\vec{\rho}}}{\theta})]+4\theta^2\tr({\vec{F}}_{i,h}{\vec{K}}_{\vec{\psi}})\\
&=-4d\theta\,\pdv{p_i}{\theta}+p_iw_i.
\end{align*}
For the product in Equation~\eqref{eq:record-product}, the Laplacian differentiates one factor twice or differentiates two distinct factors once each.
After dividing by the positive likelihood, this gives
\begin{align}
\frac{\mathcal LL_\theta}{L_\theta}
&=\sum_i\frac{\mathcal Lp_i}{p_i}+2\sum_{j<i}\frac{\nabla p_i\cdot\nabla p_j}{p_ip_j}\nonumber\\
&=-4d\theta\sum_i\pdv{}{\theta}\log p_i
+\sum_iw_i+2\sum_{j<i}\vec g_i\cdot\vec g_j\nonumber\\
&=-4d\theta\,\pdv{}{\theta}\log L_\theta+W,\label{eq:record-laplacian}
\end{align}
where
\[
W=\sum_{i=1}^Jw_i+2\sum_{j<i}\vec g_i\cdot\vec g_j.
\]
All the gradient vectors belong to the same tangent space at the fixed \({\vec{\psi}}\).

Let \(q_\theta(h)=\int L_\theta(h,{\vec{\psi}})\,\odif{\sigma}({\vec{\psi}})\) be the marginal record density relative to \(\mu\).
Multiply Equation~\eqref{eq:record-laplacian} by \(L_\theta\) and integrate over \({\vec{\psi}}\).
The integral of the left side is zero because it is a sphere Laplacian.
Differentiating under the integral, justified at the end of this proof, yields
\[
4d\theta\,\pdv{q_\theta(h)}{\theta}
=\int L_\theta(h,{\vec{\psi}})W(h,\theta,{\vec{\psi}})\,\odif{\sigma}({\vec{\psi}}).
\]
The posterior of the hidden direction satisfies
\[
\Pr_\theta(\odif{\vec\psi}\mid H=h)=\frac{L_\theta(h,\vec\psi)}{q_\theta(h)}\,\odif{\sigma}(\vec\psi).
\]
Dividing the preceding identity by \(4d\theta q_\theta(h)\) proves
\begin{equation}\label{eq:posterior-score}
\pdv{}{\theta}\log q_\theta(h)=\frac1{4d\theta}\E_\theta[W\mid H=h].
\end{equation}
This calculation uses the full likelihood inside the sphere integral.
It therefore includes the effect of all earlier observations on the posterior direction.

Fix \(\theta\) and condition on \({\vec{\psi}}\).
For complete round outcomes \(Y_1,\ldots,Y_i\), define \(\mathcal F_i:=\sigma(\omega,Y_1,\ldots,Y_i)\).
The chosen operation for round \(i\) is determined by \(\mathcal F_{i-1}\).
Lemma~\ref{lem:local} applies conditional on this history and gives
\[
\E[\vec g_i\mid\mathcal F_{i-1},{\vec{\psi}}]=0,
\qquad \E[w_i\mid\mathcal F_{i-1},{\vec{\psi}}]=0.
\]
Define \(\vec V_i=\sum_{j=1}^i\vec g_j\), with \(\vec V_0=0\).
For \(j<i\), the vector \(\vec g_j\) is measurable at time \(i-1\), so
\[
\E[\vec g_i\cdot\vec g_j\mid{\vec{\psi}}]
=\E[\vec g_j\cdot\E[\vec g_i\mid\mathcal F_{i-1},{\vec{\psi}}]\mid{\vec{\psi}}]=0.
\]
Expanding the norm squared and using Lemma~\ref{lem:local} gives
\begin{equation}\label{eq:V-bound}
\E[\norm{\vec V_{i-1}}^2\mid{\vec{\psi}}]
=\sum_{j<i}\E[\norm{\vec g_j}^2\mid{\vec{\psi}}]\le Cd(i-1).
\end{equation}

Write \(\xi_i=w_i+2\vec g_i\cdot\vec V_{i-1}\).
Since \(\vec V_{i-1}\) is \(\mathcal F_{i-1}\)-measurable, \(\E[\xi_i\mid\mathcal F_{i-1},\vec\psi]=0\).
Also \(W=\sum_i\xi_i\).
The same orthogonality argument gives \(\E[W^2\mid{\vec{\psi}}]=\sum_i\E[\xi_i^2\mid{\vec{\psi}}]\).
The elementary inequality \((a+b)^2\le2a^2+2b^2\) implies
\[
\E[\xi_i^2\mid{\vec{\psi}}]\le2\E[w_i^2\mid{\vec{\psi}}]+8\E[(\vec g_i\cdot\vec V_{i-1})^2\mid{\vec{\psi}}].
\]
Conditional on \(\mathcal F_{i-1}\) and \({\vec{\psi}}\), the covariance inequality in Lemma~\ref{lem:local} gives
\begin{align*}
\E[(\vec g_i\cdot\vec V_{i-1})^2\mid\mathcal F_{i-1},{\vec{\psi}}]
&=\vec V_{i-1}^{\mathsf T}\E[\vec g_i\vec g_i^{\mathsf T}\mid\mathcal F_{i-1},{\vec{\psi}}]\vec V_{i-1}\\
&\le C\norm{\vec V_{i-1}}^2.
\end{align*}
Taking expectations and summing, Equation~\eqref{eq:V-bound} yields
\begin{align}
\E[W^2\mid{\vec{\psi}}]
&\le CdrJ+C\sum_{i=1}^J\E[\norm{\vec V_{i-1}}^2\mid{\vec{\psi}}]\nonumber\\
&\le CdrJ+Cd\sum_{i=1}^J(i-1)
\le Cd(Jr+J^2).\label{eq:W-bound}
\end{align}
These inequalities allow arbitrary dependence between \(w_i\) and \(\vec g_i\) within a round.

Apply conditional Jensen to Equation~\eqref{eq:posterior-score}, and then average Equation~\eqref{eq:W-bound} over \({\vec{\psi}}\).
The joint expectation here includes the direction and all measurement outcomes.
We obtain
\begin{align*}
I(Q_\theta)
&=\frac1{16d^2\theta^2}\E_\theta\ab[\ab(\E_\theta[W\mid H])^2]\\
&\le\frac1{16d^2\theta^2}\E_\theta[W^2]
\le C\frac{J^2+Jr}{d}.
\end{align*}
The last inequality uses the fixed positive lower bound on \(\theta\).

The state and its derivative are diagonal in the decomposition into \({\vec{\psi}}\) and \({\vec{\psi}}^\perp\).
On these two subspaces, the ratios of derivative eigenvalues to state eigenvalues are respectively
\[
\frac{1-1/d}{\theta+(1-\theta)/d}
\quad\text{and}\quad
-\frac1{1-\theta}.
\]
For \(\theta\in[1/3,2/3]\), \(\lvert(1-1/d)/(\theta+(1-\theta)/d)\rvert\le3\) and \(1/(1-\theta)\le3\).
Thus
\[
-3{\vec{\rho}}\preceq\pdv{{\vec{\rho}}}{\theta}\preceq3{\vec{\rho}}.
\]
Taking the tensor product with a positive \({\vec{\rho}}\) preserves this operator order.
Adding the two differentiated factors gives
\[
-6{\vec{\rho}}^{\otimes2}\preceq\pdv{{\vec{\rho}}^{\otimes2}}{\theta}\preceq6{\vec{\rho}}^{\otimes2}.
\]
Applying any positive effect gives \(\lvert\pdv{p_i}{\theta}\rvert\le6p_i\), so \(s_i:=\pdv{}{\theta}\log p_i\) satisfies \(|s_i|\le6\).
Also \(\E_\theta[s_i\mid\mathcal F_{i-1},\vec\psi]=0\), since integrating the density derivative against the reference kernel differentiates its total mass of one.
By martingale orthogonality,
\[
\E_\theta\ab[\ab(\sum_{i=1}^J s_i)^2]\le36J.
\]
Differentiating the marginal likelihood expresses its score as
\[
\pdv{}{\theta}\log q_\theta(H)
=\E_\theta\ab[\sum_{i=1}^J s_i\mid H].
\]
Conditional Jensen gives \(I(Q_\theta)\le36J\).
Together with the preceding bound, this proves Lemma~\ref{lem:information}.

We now justify the density manipulations used above for arbitrary measurable outcomes.
The model assumes standard Borel classical outcome spaces and measurable policy maps.
This includes discrete outcomes, real outcome vectors, and classical descriptions of randomly chosen measurements.
The assumption permits regular conditional probabilities and measurable control using the history.

For a fixed history \(h\), the complete round has a POVM effect measure \({\vec{F}}_h\) on the two input systems.
Its outcome distribution on the maximally mixed inputs is the probability measure
\[
\nu_h(A)=\frac{\tr({\vec{F}}_h(A))}{d^2}.
\]
If \(\nu_h(A)=0\), positivity implies \({\vec{F}}_h(A)=0\), so every input state's outcome measure is absolutely continuous with respect to \(\nu_h\).
Finite dimensional operator valued Radon Nikodym derivatives give a positive density \({\vec{F}}_h(y)\) with
\[
\int {\vec{F}}_h(y)\,\odif{\nu_h}(y)={\vec{I}},
\qquad \tr({\vec{F}}_h(y))=d^2\quad\text{for }\nu_h\text{ almost every }y.
\]
The conditional likelihood ratio relative to \(\nu_h\) is \(p_h(y,\theta,{\vec{\psi}})=\tr({\vec{F}}_h(y){\vec{\rho}}_{\theta,{\vec{\psi}}}^{\otimes2})\).
These densities can be chosen jointly measurable in \(h,y\).
To verify this directly, take increasing finite partitions that generate the outcome sigma algebra, and let \(A_m(y)\) be the cell containing \(y\) in partition \(m\).
The matrices \({\vec{F}}_h(A_m(y))/\nu_h(A_m(y))\), with zero assigned when the denominator is zero, are jointly measurable.
For each fixed \(h\), they form a bounded martingale, entry by entry, under \(\nu_h\).
Their almost sure limit is the required density.
Positivity and trace \(d^2\) pass to the limit, and choosing the identity matrix on the measurable set where convergence fails gives a jointly measurable version everywhere.
On the chosen parameter interval, \({\vec{I}}/(3d)\preceq{\vec{\rho}}_{\theta,{\vec{\psi}}}\preceq {\vec{I}}\).
Consequently
\begin{equation}\label{eq:density-floor}
3^{-2}\le p_h(y,\theta,{\vec{\psi}})\le d^2.
\end{equation}
These bounds are uniform in the history and every nonzero outcome density.

Use the seed law and these reference kernels successively to define a reference measure \(\mu\) on complete records.
Equivalently, \(\mu=\operatorname{Law}_{\vec I/d}(H)\) for the same policy.
The product of conditional densities in Equation~\eqref{eq:record-product} is its likelihood ratio, and Equation~\eqref{eq:density-floor} gives
\[
3^{-2J}\le L_\theta(h,{\vec{\psi}})\le d^{2J}.
\]
The radial score calculation above also gives \(|\partial_\theta\log L_\theta|\le6J\).
For fixed \(d,J\), the sphere derivatives satisfy \(\sup_{h,\theta,\vec\psi}\norm{\nabla^aL_\theta(h,\vec\psi)}\le C_{d,J}\) for \(a\in\{0,1,2\}\).
To see this, the input projector is a polynomial in the real sphere coordinates, its derivatives are bounded on the compact sphere, and \(\norm{{\vec{F}}_h(y)}\le\tr({\vec{F}}_h(y))=d^2\).
Each local derivative is therefore bounded uniformly in \(h,y\).
The positive lower bound in Equation~\eqref{eq:density-floor} bounds derivatives of its logarithm, and the product rule gives finite uniform bounds for the complete likelihood.

These estimates justify exchanging parameter derivatives with the sphere integral by dominated convergence.
They also justify sphere integration by parts and all outcome expectations.
The density \(q_\theta\) inherits a positive lower bound and bounded derivatives, so \(\sqrt{q_\theta}\) is differentiable as an \(L^2(\mu)\) function.
This regularity justifies the testing argument in Appendix~\ref{app:one-sided-test}.

Finally, the first instrument also admits the refinement used in Lemma~\ref{lem:local} for general outcomes.
Associate to it its finite dimensional positive Choi operator measure.
Let \(\nu_{\mathrm M}\) be its trace measure, which dominates every matrix entry.
The resulting positive Choi density has a measurable positive square root.
Reshaping its columns into maps from \(\CC^d\) to \(\CC^r\) gives finitely many measurable Kraus maps for each outcome.
Their completeness relation is \(\sum_a\int\vec M_{s,a}^\dagger\vec M_{s,a}\,\odif{\nu_{\mathrm M}}(s)=\vec I\).
The original second measurement and later policy continue to ignore the extra column label.
The Cauchy inequality in Equation~\eqref{eq:merge} applies to these densities, so the local bounds remain valid after summing or integrating out that label.
This completes the justification for general outcome spaces.
\end{proof}

\subsection{Proof of Remark~\ref{rem:information-sharpness}}\label{app:information-sharpness}

\informationsharpness*

More precisely, for integers \(J\ge1\), \(n\ge1\), and \(0\le k\le n\), let \(d=2^n\) and \(r=2^k\).
For the state family in Equation~\eqref{eq:states}, there is a protocol \(\mathcal A_{J,d,r}\) using exactly \(J\) rounds in the reset memory model, whose operations depend only on \(J,d,r\), such that
\[
I(Q_\theta^{\mathcal A_{J,d,r}})\ge c\min\ab\{J,\frac{J^2+Jr}{d}\}\qquad\text{for every }\theta\in[1/3,2/3].
\]
Consequently, with universal implicit constants,
\[
\sup_{\mathcal A:\,J\text{ rounds}}I(Q_\theta^{\mathcal A})=\Theta\ab(\min\ab\{J,\frac{J^2+Jr}{d}\}).
\]

\begin{proof}
For a bounded statistic \(Y\) computed from the complete record, differentiation of its expectation and the Cauchy inequality give
\[
\ab(\pdv{}{\theta}\E_\theta Y)^2
=\ab(\E_\theta[(Y-\E_\theta Y)\pdv{}{\theta}\log q_\theta(H)])^2
\le\Var_\theta(Y)I(Q_\theta).
\]
The regularity established for Lemma~\ref{lem:information} justifies this differentiation.
Here \(P_d'(\theta)=2(1-1/d)\theta\ge1/3\), so a bounded unbiased purity statistic satisfies \(I(Q_\theta)\ge1/[9\Var_\theta(Y)]\).

First use no quantum memory.
For \(J\le d\), measure all \(2J\) copies in one independently sampled Haar basis and set
\[
Y=\frac{d+1}{2J(2J-1)}\sum_{u\ne v}\ind\{X_u=X_v\}-1.
\]
For each fixed direction, Equation~\eqref{eq:collision-variance} with one batch gives the same mean \(P_d(\theta)\) and a uniform variance bound.
Averaging over the direction therefore gives
\[
\E_\theta Y=P_d(\theta),\qquad
\Var_\theta(Y)\le C\ab(\frac1d+\frac1J+\frac d{J^2})\le C\frac d{J^2}.
\]
Thus \(I(Q_\theta)\ge cJ^2/d\).
For \(J>d\), take \(B=\lfloor J/d\rfloor\) batches of \(d\) pairs each, with independent Haar bases, and average their collision statistics \(Y_1,\ldots,Y_B\).
Discard the remaining pairs to obtain exactly \(J\) rounds.
Conditional on the single hidden direction \(\vec\psi\), the batches are independent and satisfy
\[
\E_\theta[Y_b\mid\vec\psi]=P_d(\theta),\qquad
\Var_\theta(Y_b\mid\vec\psi)\le C/d.
\]
The conditional mean is independent of the direction.
The law of total variance therefore yields
\[
\Var_\theta\ab(\frac1B\sum_{b=1}^B Y_b)\le\frac{C}{Bd}\le\frac{2C}{J},
\]
where \(B\ge J/(2d)\).
This gives \(I(Q_\theta)\ge cJ\).
The two collision constructions achieve \(cA_J\), where \(A_J:=\min\{J,J^2/d\}\).
The constant conditional mean is what allows the variance reduction despite the shared direction.

For \(r\ge2\), use the block experiment from Equation~\eqref{eq:block-law} independently in every round.
After averaging the fresh rotation in each pair, its antisymmetric indicator has conditional probability
\[
\Pr_\theta(A_t=1\mid\vec\psi)=\kappa_r[1-P_d(\theta)],\qquad
\kappa_r=\frac{r-1}{2(d-1)}.
\]
This probability does not depend on \(\vec\psi\), so the indicators remain independent Bernoulli variables after averaging that direction.
Their Fisher information is exactly
\[
I_\theta(A_1,\ldots,A_J)
=J\kappa_r\frac{[P_d'(\theta)]^2}
{[1-P_d(\theta)][1-\kappa_r(1-P_d(\theta))]}
=\Theta(Jr/d).
\]
Indeed, \(P_d'(\theta)\in[1/3,4/3]\), \(1-P_d(\theta)\in[5/18,8/9]\), \(\kappa_r\le1/2\), and \(\kappa_r=\Theta(r/d)\) uniformly for \(2\le r\le d\).
The indicators are functions of the complete record, so Fisher information data processing gives \(I(Q_\theta)\ge I_\theta(A_1,\ldots,A_J)\).

Set \(M_J:=Jr/d\).
Since \(M_J\le J\),
\[
\max\{A_J,M_J\}\le\min\ab\{J,\frac{J^2+Jr}{d}\}
\le A_J+M_J\le2\max\{A_J,M_J\}.
\]
If \(r=1\), then \(M_J\le A_J\), since \(J\ge1\).
Choose collisions when \(J\ge r\), and otherwise choose the block experiment.
This selects a construction achieving the larger scale, using only \(J,d,r\).
The lower bound follows, and Lemma~\ref{lem:information} gives the matching upper bound.
\end{proof}

\subsection{Proof of Lemma~\ref{lem:one-sided-test}}
\label{app:one-sided-test}

\begin{lemma}[Testing with a one-sided cost bound]
\label{lem:one-sided-test}
There are universal \(c,h_0>0\) such that the following holds for \(d=2^n\), \(r=2^k\), integers \(n\ge1\), \(0\le k\le n\), and \(1/3\le\theta_0<\theta_1\le2/3\) with \(h:=\theta_1-\theta_0\le h_0\).
A tester receives copies of \(\vec\rho_{\theta,\vec\psi}\) from Equation~\eqref{eq:states}, where one Haar direction \(\vec\psi\) is drawn and fixed, and processes copies in pairs with retained dimension at most \(r\) and a reset after each pair.
At each pair boundary, including before the first pair, it returns an answer or selects its next measurement using the classical history and state-independent randomness.
If it terminates almost surely at both endpoints and its output \(Z\in\{0,1\}\) satisfies \(\Pr_{\theta_0}(Z=1)\le\delta\) and \(\Pr_{\theta_1}(Z=0)\le\delta\) for \(0<\delta\le1/16\), then its number \(T_{\mathrm{test}}\) of copies from the tested source satisfies
\[
\E_{\theta_0}T_{\mathrm{test}}\ge c\Rrate{h}.
\]
The probabilities and expectation average over the hidden direction and the tester's randomness.
\end{lemma}

\begin{proof}
Write \(m=\E_{\theta_0}T_{\mathrm{test}}\).
If \(m=+\infty\), the claimed bound is immediate.
Otherwise define the deterministic cap on pairs from the tested source
\[
J=\max\{1,\lceil8m\rceil\}.
\]
This cap may depend on the base law, but the capped tester uses the same value of \(J\) at every parameter.

At each pair boundary, first process the original stopping decision.
If the tester requests another pair after \(J\) pairs have been completed, declare \(\theta_1\).
Let \(\widetilde Z\in\{0,1\}\) be this capped answer, with \(1\) denoting \(\theta_1\).
Couple the capped and original testers using the same measurement outcomes and random seeds until the cap occurs.
The cap is reached only when the original tester would use more than \(2J\) copies.
Markov's inequality and \(2J\ge16m\) therefore give
\[
\Pr_{\theta_0}(T_{\mathrm{test}}>2J)\le\frac m{2J}\le\frac1{16}.
\]
Consequently the endpoint errors of the capped test obey
\begin{equation}
\label{eq:one-sided-errors}
\Pr_{\theta_0}(\widetilde Z=1)\le\delta+\frac1{16},
\qquad \Pr_{\theta_1}(\widetilde Z=0)\le\delta.
\end{equation}
The second inequality holds because a capped answer of \(0\) must be an original answer of \(0\).
For \(\delta\le1/16\),
\begin{equation}
\label{eq:one-sided-separation}
\TV(P_{\theta_0}^{\widetilde Z},P_{\theta_1}^{\widetilde Z})\ge1-2\delta-\frac1{16}\ge\frac34.
\end{equation}

After stopping, fill any remaining rounds with discarded fresh pairs, without changing \(\widetilde Z\).
The resulting experiment is an admissible fixed-budget protocol with exactly \(J\) rounds and retained register dimension at most \(r\).
Include the state-independent seed and the capped answer in its complete record.

Let \(Q_\theta\) denote its complete record law, with density \(q_\theta\) under the common reference measure from Appendix~\ref{sec:regularity}.
That appendix shows that \(\sqrt{q_\theta}\) is differentiable in \(L^2\) and
\[
\norm{\pdv{\sqrt{q_\theta}}{\theta}}_2=\frac12\sqrt{I(Q_\theta)}.
\]
With \(h=\theta_1-\theta_0\), the fundamental theorem of calculus, the triangle inequality in \(L^2\), and output data processing give
\begin{align}
\TV(P_{\theta_0}^{\widetilde Z},P_{\theta_1}^{\widetilde Z})
&\le\TV(Q_{\theta_0},Q_{\theta_1})\le\norm{\sqrt{q_{\theta_1}}-\sqrt{q_{\theta_0}}}_2\nonumber\\
&\le\frac12\int_{\theta_0}^{\theta_1}\sqrt{I(Q_\theta)}\,\odif{\theta}\nonumber\\
&\le C h\sqrt{\min\ab\{J,\frac{J^2+Jr}{d}\}}.\label{eq:one-sided-information}
\end{align}
For the first line, factor \(q_{\theta_1}-q_{\theta_0}\) as a product of the difference and sum of their square roots and use that the sum has \(L^2\) norm at most \(2\).
The last line is Lemma~\ref{lem:information}, uniform over the entire interpolation interval.
The test is defined at all intermediate parameters, although its error and cost assumptions are used only at the endpoints.

Equations~\eqref{eq:one-sided-separation} and~\eqref{eq:one-sided-information} imply universal \(c_1,c_2>0\) such that
\[
J\ge c_1h^{-2},\qquad J^2+Jr\ge c_2d/h^2.
\]
The second inequality gives either \(J^2\ge c_2d/(2h^2)\) or \(Jr\ge c_2d/(2h^2)\).
Hence
\[
J\ge c_3\max\ab\{h^{-2},\min\ab\{\frac{\sqrt d}{h},\frac d{rh^2}\}\}=c_3\Rrate{h}.
\]
Choose the universal \(h_0>0\) small enough that \(c_1h_0^{-2}\ge2\).
Then \(J\ge2\), whereas \(J\le8m+1\), so
\[
m\ge\frac{J-1}{8}\ge\frac J{16}\ge c\Rrate{h}.
\]
This proves the lemma using only the base expected cost for the tested source.
\end{proof}

\subsection{Proof of Theorem~\ref{thm:estimation-lower}}\label{sec:testing-details}

\begin{theorem}[Purity estimation lower bound]\label{thm:estimation-lower}
There exist universal constants \(c,\eps_{\mathrm L}>0\) such that, for \(d=2^n\), \(r=2^k\), integers \(n\ge1\) and \(0\le k\le n\), and \(0<\eps\le\eps_{\mathrm L}\), every almost surely terminating estimator \(A\) in the model of Section~\ref{sec:model} satisfying \(\Pr_{\vec\rho}(|\widehat P-P(\vec\rho)|\le\eps)\ge3/4\) for every state \(\vec\rho\) obeys
\[
\sup_{\vec\rho}\E_{\vec\rho}T_A
\ge\E_{\vec\psi\sim\sigma}\E_{\vec\rho_{1/2,\vec\psi}}T_A
\ge c\max\ab\{\eps^{-2},\min\ab\{\sqrt d/\eps,d/(r\eps^2)\}\}.
\]
Here \(\sigma\) is Haar probability measure, and the baseline parameter \(1/2\) and prior do not depend on the estimator or accuracy.
Classical adaptation and random stopping times are unrestricted.
\end{theorem}

\begin{proof}
For an admissible estimator \(A\), put
\[
m:=\E_{\vec\psi\sim\sigma}\E_{\vec\rho_{1/2,\vec\psi}}T_A.
\]
If \(m=+\infty\), the prior-average lower bound and its worst-case consequence are immediate.
Suppose \(m<\infty\).
Set \(\theta_0=1/2\), \(\theta_1=1/2+4\eps\), and choose \(\eps_{\mathrm L}\le\min\{1/24,h_0/4\}\), where \(h_0\) is the constant in Lemma~\ref{lem:one-sided-test}.
Then \(\theta_0,\theta_1\in[1/3,2/3]\) and their separation is at most \(h_0\).
Equation~\eqref{eq:purity-family} gives
\begin{align*}
P_d(\theta_1)-P_d(\theta_0)
&=\ab(1-\frac1d)\ab(\ab(\tfrac12+4\eps)^2-\tfrac14)\\
&=\ab(1-\frac1d)(4\eps+16\eps^2)>2\eps.
\end{align*}
This strict separation holds also at \(d=2\), where its right side before the inequality is \(2\eps+8\eps^2\).

Draw one Haar direction \(\vec\psi\) and keep it fixed for the entire test, including every repetition.
For each endpoint \(\theta_b\), run \(23\) repetitions of \(A\) sequentially on fresh copies of the same state \(\vec\rho_{\theta_b,\vec\psi}\), obtaining outputs \(\widehat P_1,\ldots,\widehat P_{23}\).
Use independent algorithm seeds and disjoint fresh-copy streams for the runs.
Conditional on the shared fixed direction \(\vec\psi\), these outputs and their associated stopping times are independent.
Each run terminates almost surely and its error indicator has conditional expectation at most \(1/4\).
For the median \(\widehat P_{\mathrm{med}}=\operatorname{median}(\widehat P_1,\ldots,\widehat P_{23})\), an error larger than \(\eps\) requires at least \(12\) erroneous runs.
Hoeffding's inequality, conditional on the fixed direction, yields
\begin{align*}
\Pr_{\theta_b,\vec\psi}(|\widehat P_{\mathrm{med}}-P_d(\theta_b)|>\eps)
&\le\Pr_{\theta_b,\vec\psi}\ab(\sum_{\ell=1}^{23}\ind\{|\widehat P_\ell-P_d(\theta_b)|>\eps\}\ge23/2)\\
&\le\exp\ab(-2\cdot23\ab(\frac12-\frac14)^2)
=\exp(-23/8)<\frac1{16}.
\end{align*}
The bound is uniform in \(\vec\psi\), so integration gives the same endpoint error bound under Haar averaging.

If a run stops after the first copy of a pair, keep its classical answer, discard any retained quantum register, and request and discard one unused second copy.
Otherwise no copy is added.
For its original copy cost \(T_{A,\ell}\), the resulting complete-pair cost is
\[
S_\ell=2\lceil T_{A,\ell}/2\rceil\le T_{A,\ell}+1.
\]
In particular, a run stopping before any query has \(S_\ell=0\).
Any zero-copy runs are incorporated into the classical decision at the current pair boundary.
There are at most \(23\) runs, so these decisions preserve almost sure termination.
Every run starts with an empty register, so no quantum system survives from one run to the next.
Padding changes no estimator answer or error probability.

Define the tester by
\[
Z=\ind\ab\{\widehat P_{\mathrm{med}}>\frac{P_d(\theta_0)+P_d(\theta_1)}2\ab\},
\qquad T_{\mathrm{test}}=\sum_{\ell=1}^{23}S_\ell.
\]
Because the two purities differ by more than \(2\eps\), an accurate median produces the correct endpoint answer.
The test therefore has both errors at most \(1/16\), processes copies in complete reset pairs, and terminates almost surely under both hypotheses.
The random seeds have no dependence on the input parameter or direction.
Its base expected cost satisfies
\[
\E_{\theta_0}T_{\mathrm{test}}\le23(m+1).
\]
This averaged bound follows by applying linearity of expectation conditional on the shared fixed direction and then integrating over its Haar law.

Lemma~\ref{lem:one-sided-test}, applied with separation \(4\eps\), now gives a universal \(c_{\mathrm t}>0\) such that
\[
23(m+1)\ge c_{\mathrm t}\Rrate{4\eps}\ge\frac{c_{\mathrm t}}{16}\Rrate{\eps}.
\]
For the last inequality, the inverse-square terms decrease by \(1/16\) and the inverse-linear term by \(1/4\), so taking their minimum and maximum preserves a factor of at least \(1/16\).
Write \(a_0=c_{\mathrm t}/(16\cdot23)\) and decrease \(\eps_{\mathrm L}\) further so that \(\eps_{\mathrm L}\le\sqrt{a_0/2}\).
Since \(\Rrate{\eps}\ge\eps^{-2}\), this makes \(a_0\Rrate{\eps}\ge2\) and hence
\[
m\ge a_0\Rrate{\eps}-1\ge\frac{a_0}{2}\Rrate{\eps}.
\]
The choice of \(\eps_{\mathrm L}\) absorbs the additive padding cost, including on zero-copy branches of the original estimator.
Finally, \(\sup_{\vec\rho}\E_{\vec\rho}T_A\ge m\) because the prior is supported on valid states.
\end{proof}

\section{Purest State Identification: Upper Bounds}\label{sec:identification-upper}

\subsection{Proof of Lemma~\ref{lem:estimate-primitive}}
\label{app:estimate-primitive}

\begin{lemma}[Purity estimation primitive]
\label{lem:estimate-primitive}
For \(d=2^n\ge2\), \(r=2^k\) with \(0\le k\le n\), \(0<a\le1\), and \(0<\eta<1/2\), \(\textnormal{\textsc{Estimate}}(i,a,\eta)\) uses a deterministic even budget of at most \(C\Rrate{a}\log(2/\eta)\) copies and returns an estimate satisfying
\[
\Pr_\nu(|\widehat\lambda_i-\lambda_i|>a)\le\eta.
\]
Each call uses fresh copies and independent randomness.
Each call uses one \(k\)-qubit register and resets it after every pair.
\end{lemma}

\begin{proof}
We specify the two constant-success experiments before choosing the cheaper one for the requested accuracy.
All budgets below depend on \(d,r,a,\eta\), and not on the unknown purity.
Write \(\PhiCost{a}:=\max\{a^{-2},\sqrt d/a\}\) for the collision rate.
Fix an instance \(\nu\), and let \(\mathcal H_0\) be the sigma-field generated by the classical history before the call.
The requested source \(i\), accuracy \(a\), and confidence level \(\eta\) may be \(\mathcal H_0\)-measurable.
At the start of each call, the experiment uses fresh copies of the selected state and classical random seeds independent of that history.
It therefore suffices to prove the following bounds for each fixed state and requested parameters, since the same bounds then hold conditional on that history.

In the collision experiment, let \(B\ge1\) and even \(m\ge2\) be integers.
Each batch \(b\in\{1,\ldots,B\}\) draws one Haar unitary \(\vec U_b\) and measures \(m\) fresh copies in that same basis, obtaining \(X_{b,1},\ldots,X_{b,m}\).
Different batches use independent bases and disjoint copies.
The batch statistic and the returned average are
\[
Y_b=\frac{d+1}{m(m-1)}\sum_{u\ne v}\ind\{X_{b,u}=X_{b,v}\}-1,\qquad \widehat\lambda_{\mathrm{col}}=\frac1B\sum_{b=1}^B Y_b.
\]
Conditional on \(\vec U_b\), let \(q_x=\bra x\vec U_b\vec\rho_i\vec U_b^\dagger\ket x\) for \(x\in\{1,\ldots,d\}\).
Let \(\vec S\) swap the two tensor factors of \(\CC^d\otimes\CC^d\).
The outcomes are independent with probabilities \(q_x\).
The Haar second-moment identity is
\[
\E_{\vec U}\sum_x(\vec U^\dagger\ket x\bra x\vec U)^{\otimes2}=\frac{\vec I+\vec S}{d+1}.
\]
Taking its trace against \(\vec\rho_i^{\otimes2}\) gives
\[
\E[Y_b\mid\vec U_b]=(d+1)\sum_xq_x^2-1,\qquad \E Y_b=\lambda_i.
\]
Thus unbiasedness averages both the random basis and its measurement outcomes.

We justify applying the known cross-collision variance bound of \citet{all} to this statistic.
For even \(m\), independently of the outcomes draw a uniform ordered partition into disjoint index sets \(\mathcal I,\mathcal J\) of size \(s=m/2\), and define
\[
\widetilde Y_b=\frac{d+1}{s^2}\sum_{u\in\mathcal I}\sum_{v\in\mathcal J}\ind\{X_{b,u}=X_{b,v}\}-1.
\]
For \(u\ne v\), the inclusion probability is \(\Pr(u\in\mathcal I,v\in\mathcal J)=s^2/[m(m-1)]\).
Consequently \(\E[\widetilde Y_b\mid\vec U_b,X_{b,1},\ldots,X_{b,m}]=Y_b\).
The two groups, conditional on the chosen partition and basis, are independent samples from the same state, so they are exactly the specialization with identical inputs of the experiment of \citet{all}.
Theorem 13 and Equation (116) of \citet{all}, together with the law of total variance, yield a universal \(C_{\mathrm v}\ge1\) such that
\[
\Var(Y_b)\le\Var(\widetilde Y_b)\le C_{\mathrm v}\ab(\frac1d+\frac d{m^2}+\frac1m).
\]
The independent batches therefore give
\begin{equation}\label{eq:collision-variance}
\Var(\widehat\lambda_{\mathrm{col}})\le\frac{C_{\mathrm v}}B\ab(\frac1d+\frac d{m^2}+\frac1m).
\end{equation}
The auxiliary partition is used only in this analysis.
The implemented statistic uses all ordered pairs of distinct observations, or equivalently the batch histogram.

Set \(c_0=c_1=12C_{\mathrm v}\) and choose the deterministic integers
\begin{align*}
B(a)&=\lceil c_0\max\ab\{1,1/(da^2)\}\rceil,\\
m(a)&=2\lceil(c_1/2)\max\ab\{2,\sqrt{d/(B(a)a^2)},1/(B(a)a^2)\}\rceil.
\end{align*}
Here \(\lceil x\rceil\) denotes the least integer at least \(x\).
In particular, \(m(a)\ge2\) is even, so every batch consists of complete reset pairs.
The three variance contributions obey
\[
\Var(\widehat\lambda_{\mathrm{col}})\le C_{\mathrm v}a^2(c_0^{-1}+c_1^{-2}+c_1^{-1})\le a^2/4.
\]
Chebyshev's inequality therefore gives \(\Pr(|\widehat\lambda_{\mathrm{col}}-\lambda_i|>a)\le1/4\).
The two accuracy regimes give
\[
(B(a),m(a))=\begin{cases}
(\Theta((da^2)^{-1}),\Theta(d)),&0<a\le d^{-1/2},\\
(\Theta(1),\Theta(\sqrt d/a)),&d^{-1/2}<a\le1.
\end{cases}
\]
All constants here are universal because \(c_0,c_1\) are fixed.
Thus the collision budget \(T_{\mathrm{col}}(a)=B(a)m(a)\) is a deterministic even integer with \(T_{\mathrm{col}}(a)\le C\PhiCost{a}\).

For \(r\ge2\), fix an orthogonal decomposition into \(d/r\) blocks of dimension \(r\), with projections satisfying
\[
\vec P_\ell^0\vec P_h^0=\ind\{\ell=h\}\vec P_\ell^0,\qquad \sum_{\ell=1}^{d/r}\vec P_\ell^0=\vec I,\qquad \rank(\vec P_\ell^0)=r.
\]
The assumed dimensions \(d=2^n\), \(r=2^k\), and \(k\le n\) ensure that \(d/r\) is an integer.
For every pair draw a fresh Haar unitary \(\vec U\) and use the rotated projections \(\vec P_\ell(\vec U)=\vec U\vec P_\ell^0\vec U^\dagger\).
Measure the first copy's block label and retain its postmeasurement state isometrically in the \(r\)-dimensional register.
Measure the second copy in the same decomposition.
For unequal labels record \(A_t=0\).
For equal labels perform the SWAP measurement on the two block states and record \(A_t=1\) only for its antisymmetric outcome.
All pairs, including those with unequal labels, consume two copies and reset the register.
Define the full antisymmetric projector by \(\vec\Pi_-=(\vec I-\vec S)/2\).
For a fixed rotation, the effect of the event \(A_t=1\) on the original pair is
\[
\vec Q_{\vec U}^-=\sum_{\ell=1}^{d/r}(\vec P_\ell(\vec U)\otimes\vec P_\ell(\vec U))\vec\Pi_-.
\]
Each summand projects onto the antisymmetric subspace within one block, and these subspaces are mutually orthogonal.
Their support and traces give
\[
\vec Q_{\vec U}^-=\vec\Pi_-\vec Q_{\vec U}^-\vec\Pi_-,\qquad \tr(\vec Q_{\vec U}^-)=\frac dr\frac{r(r-1)}2=\frac{d(r-1)}2,\qquad \tr(\vec\Pi_-)=\frac{d(d-1)}2.
\]
The Haar average commutes with \(\vec V\otimes\vec V\) for every unitary \(\vec V\), so the second unitary twirl lies in \(\operatorname{span}\{\vec I,\vec S\}\).
Antisymmetric support forces this average to be a scalar multiple of \(\vec\Pi_-\), and preserving its trace determines the scalar:
\[
\E_{\vec U}\vec Q_{\vec U}^-=\frac{r-1}{d-1}\vec\Pi_-.
\]
The identity \(\tr[\vec S(\vec\rho_i\otimes\vec\rho_i)]=\tr(\vec\rho_i^2)=\lambda_i\) follows by expansion in an orthonormal basis.
Taking the trace of the averaged effect against \(\vec\rho_i^{\otimes2}\) therefore gives the exact law
\begin{equation}\label{eq:block-law}
A_t\sim\operatorname{Bernoulli}(\kappa_r[1-\lambda_i]),\qquad \kappa_r=\frac{r-1}{2(d-1)}.
\end{equation}
Fresh copies and independently drawn rotations make the outputs independent across \(t\).
Choose
\[
J_r(a)=\lceil4/(\kappa_r a^2)\rceil,\qquad \widehat\lambda_{\mathrm{mem}}=1-\frac1{\kappa_rJ_r(a)}\sum_{t=1}^{J_r(a)}A_t.
\]
Then \(\E\widehat\lambda_{\mathrm{mem}}=\lambda_i\) and \(\Var(\widehat\lambda_{\mathrm{mem}})\le1/[J_r(a)\kappa_r]\le a^2/4\), so Chebyshev again gives failure probability at most \(1/4\).
Since \(\kappa_r^{-1}\le4d/r\), its deterministic even budget is \(T_{\mathrm{mem}}(a)=2J_r(a)\le Cd/(ra^2)\).
For comparison set \(T_{\mathrm{mem}}(a)=+\infty\) when \(r=1\), where no antisymmetric block event is available.

For every call and its current accuracy \(a\), compute both budgets and choose
\[
T_{\mathrm{base}}(a)=\min\{T_{\mathrm{col}}(a),T_{\mathrm{mem}}(a)\}.
\]
Their orders compare \(\max\{a^{-2},\sqrt d/a\}\) with \(d/(ra^2)\), and their minimum satisfies \(T_{\mathrm{base}}(a)\le C\Rrate{a}\).

Take an odd number of independent runs of this chosen experiment,
\[
b(\eta)=\min\{q\in\{1,3,5,\ldots\}:q\ge8\log(1/\eta)\},\qquad \widehat\lambda_i=\operatorname{median}(\widehat\lambda_i^{(1)},\ldots,\widehat\lambda_i^{(b(\eta))}).
\]
Each inaccuracy indicator has expectation at most \(1/4\), and an inaccurate median requires at least \(b(\eta)/2\) inaccurate runs.
Hoeffding's inequality gives
\[
\Pr(|\widehat\lambda_i-\lambda_i|>a)\le\exp(-b(\eta)/8)\le\eta.
\]
This calculation applies conditional on the history before the call under every fixed instance \(\nu\), because all its copies and random seeds are fresh and the error bound is uniform over the selected state.
Thus
\[
\Pr_\nu(|\widehat\lambda_i-\lambda_i|>a\mid\mathcal H_0)\le\eta\qquad\text{almost surely}.
\]
The number of runs obeys \(b(\eta)<8\log(1/\eta)+2\), so
\begin{equation}
\label{eq:id-primitive}
b(\eta)T_{\mathrm{base}}(a)\le C\Rrate{a}\log(2/\eta).
\end{equation}
This budget is deterministic and even.
Both experiments reset the register after every pair and retain only classical measurement data between pairs.
\end{proof}

\subsection{Proof of Theorem~\ref{thm:id-upper}}
\label{app:id-upper}

\identificationupper*

\begin{proof}
Fix an instance \(\nu\), and write \(S_j\) for the active set at the start of each executed epoch \(j\).
Index the estimates and interval endpoints by \((i,j)\) when their calls are requested.
Let \(\mathcal E\) be the event that every requested estimate is within \(a_j\) of its source purity.
Conditional on the history preceding any requested call, Lemma~\ref{lem:estimate-primitive} bounds its error probability by \(\delta/(2Kj^2)\).
Thus the event that source \(i\)'s call at epoch \(j\) is requested and inaccurate has probability at most \(\delta/(2Kj^2)\).
A union bound over the countably many source and epoch pairs gives
\[
\Pr_\nu(\mathcal E^c)\le\sum_{j\ge1}\sum_{i=1}^K\frac{\delta}{2Kj^2}<\delta.
\]

On \(\mathcal E\), the confidence intervals satisfy
\[
\lambda_i-2a_j\le L_{i,j}\le\lambda_i\le U_{i,j}\le\lambda_i+2a_j
\qquad(i\in S_j).
\]
In particular, \(U_{i^\star,j}\ge\lambda_{i^\star}\ge\max_{h\in S_j}L_{h,j}\), so the elimination rule retains the best source at every executed epoch.
For each challenger define
\[
j_i=\min\{j\ge1:2^{-j}<\Delta_i\},\qquad
j_{i^\star}=\max_{i\ne i^\star}j_i.
\]
At epoch \(j_i\), any still-active challenger \(i\) satisfies
\[
U_{i,j_i}\le\lambda_i+2a_{j_i}<\lambda_{i^\star}-2a_{j_i}\le L_{i^\star,j_i},
\]
because \(4a_{j_i}=2^{-j_i}<\Delta_i\).
The elimination rule removes it at this epoch.
Consequently, on \(\mathcal E\), the algorithm returns \(i^\star\) by epoch \(j_{i^\star}\), and source \(i\) is sampled through at most epoch \(j_i\).

The rate satisfies
\begin{equation}
\label{eq:id-growth}
2\Rrate{a}\le\Rrate{a/2}\le4\Rrate{a}.
\end{equation}
Indeed, the terms \(a^{-2}\), \(\sqrt d/a\), and \(d/(ra^2)\) grow by factors \(4\), \(2\), and \(4\), respectively, and taking their minimum and maximum preserves these bounds.
Since every gap is at most \(1\), the first crossing obeys \(\bar\Delta_i/8\le a_{j_i}<\bar\Delta_i/4\), including \(j_i=1\).
Equation~\eqref{eq:id-primitive} bounds source \(i\)'s total copy cost on \(\mathcal E\) by
\[
C\sum_{j=1}^{j_i}\Rrate{a_j}\log\frac{4Kj^2}{\delta}
\le 2C\Rrate{a_{j_i}}\log\frac{4Kj_i^2}{\delta}.
\]
Here the logarithm is increasing in \(j\), and Equation~\eqref{eq:id-growth} gives \(\Rrate{a_j}\le2^{j-j_i}\Rrate{a_{j_i}}\), so the sum is geometric.
Moreover, \(j_i\le C\log(e/\bar\Delta_i)\) and \(\Rrate{a_{j_i}}\le C\Rrate{\bar\Delta_i}\).
Summing over all sources therefore gives, on \(\mathcal E\),
\begin{equation}
\label{eq:id-good-cost}
T\le C\sum_{i=1}^K\Rrate{\bar\Delta_i}\ab[\log\frac K\delta+\log\log\frac e{\bar\Delta_i}\ab].
\end{equation}
Both this bound and \(\widehat i=i^\star\) hold on \(\mathcal E\), whose probability is at least \(1-\delta\).
This proves Equation~\eqref{eq:id-upper} and the stated correctness guarantee.
Each estimator uses the same \(k\)-qubit register and resets it after every pair.
\end{proof}

\section{Purest State Identification: Lower Bounds}\label{sec:identification-lower}

\subsection{Proof of Lemma~\ref{lem:id-transportation}}
\label{app:id-transportation}

Quantum relative entropy charges the confidence cost to the source whose state changes.
For density matrices \(\vec\rho,\vec\sigma\), define
\[
D(\vec\rho\Vert\vec\sigma)=\begin{cases}
\tr[\vec\rho(\log\vec\rho-\log\vec\sigma)],&\operatorname{supp}(\vec\rho)\subseteq\operatorname{supp}(\vec\sigma),\\
+\infty,&\text{otherwise}.
\end{cases}
\]
For probability laws \(P,Q\), define
\[
\KL(P\Vert Q)=\begin{cases}
\displaystyle\int\log\ab(\odv{P}{Q})\,\odif{P},&P\ll Q,\\
+\infty,&\text{otherwise},
\end{cases}
\]
Here \(\odv{P}{Q}\) denotes the Radon Nikodym density, with the usual zero-probability conventions.
For a finite output set \(\mathcal O\),
\[
\KL(P\Vert Q)=\sum_{o\in\mathcal O}P(o)\log\frac{P(o)}{Q(o)}.
\]
For Bernoulli laws, write \(\kl(p,q)=p\log(p/q)+(1-p)\log((1-p)/(1-q))\), with the usual endpoint conventions.

\begin{lemma}[Sourcewise transportation in the reset model]
\label{lem:id-transportation}
Let \(\nu=(\vec\rho_i)_{i=1}^K\) and \(\nu'=(\vec\rho_i')_{i=1}^K\) be two environments.
A protocol in the same-source reset model that returns \(O\in\mathcal O\), with \(\lvert\mathcal O\rvert<\infty\), within a deterministic horizon of \(H\) rounds satisfies
\begin{equation}
\label{eq:id-transportation}
\KL(P_\nu^O\Vert P_{\nu'}^O)\le\sum_{i=1}^K\E_\nu N_i(T)D(\vec\rho_i\Vert\vec\rho_i').
\end{equation}
\end{lemma}

\begin{proof}
Include the learner's state-independent random seed in the classical record, so its choices are fixed conditional on the preceding history.
After termination, pad the record with a fixed dummy outcome through round \(H\), and write \(I_t=0\) for these rounds, which request no copies.
Let \(Z_{t-1}\) denote the history before round \(t\), including the seed and any returned answer.
For a history \(h\) selecting source \(i\ge1\), the complete pair measurement defines one quantum-to-classical channel \(\mathcal M_{t,h}\).
This channel includes the first instrument, its classical outcome, and the conditional measurement on the retained register and second copy.
The reset ensures that its input is the fresh pair \(\vec\rho_i^{\otimes2}\), with no quantum system from earlier rounds.
Write \(P_{t,h}^\nu\) and \(P_{t,h}^{\nu'}\) for its outcome laws under the two environments.
Quantum data processing \citep{lindblad1975completely} and tensor-product additivity give
\begin{align*}
\KL(P_{t,h}^\nu\Vert P_{t,h}^{\nu'})
&\le D(\vec\rho_i^{\otimes2}\Vert(\vec\rho_i')^{\otimes2})\\
&=2D(\vec\rho_i\Vert\vec\rho_i').
\end{align*}
For a dummy round, the two conditional laws coincide and their divergence is zero.
Stopping and source selection also have the same rule under both environments at each fixed history.
The seed has the same law under both environments, so the classical chain rule yields
\begin{align*}
\KL(P_\nu^{Z_H}\Vert P_{\nu'}^{Z_H})
&=\sum_{t=1}^H\E_\nu\KL(P_{t,Z_{t-1}}^\nu\Vert P_{t,Z_{t-1}}^{\nu'})\\
&\le2\sum_{t=1}^H\sum_{i=1}^K\Pr_\nu(I_t=i)D(\vec\rho_i\Vert\vec\rho_i')\\
&=\sum_{i=1}^K\E_\nu N_i(T)D(\vec\rho_i\Vert\vec\rho_i'),
\end{align*}
where \(N_i(T)=2\sum_{t=1}^H\ind\{I_t=i\}\).
For general measurable outcomes, the same finite-horizon chain rule follows by factoring the conditional record densities and integrating their logarithms under the base law.
The answer \(O\) is a function of \(Z_H\), so classical data processing gives
\[
\KL(P_\nu^O\Vert P_{\nu'}^O)\le\KL(P_\nu^{Z_H}\Vert P_{\nu'}^{Z_H}),
\]
which proves Equation~\eqref{eq:id-transportation}.
The inequality holds in the extended nonnegative reals, with a zero factor contributing zero.
\end{proof}

\subsection{Proof of Theorem~\ref{thm:id-lower}}
\label{app:id-lower}

\identificationlower*

\begin{proof}
Choose \(\delta_0\le1/16\) and \(\Delta_0\le1/32\), and let \(\underline\Delta=\min_{2\le j\le K}\Delta_j\).
The construction uses Equation~\eqref{eq:states} with one fixed direction per source throughout the interaction.
For \(\theta_0=1/2\), define separately for each challenger
\begin{equation}
\label{eq:id-parameters}
\theta_{j,\pm}=\sqrt{\frac14\pm\frac{\Delta_j}{1-1/d}},\qquad j\in\{2,\ldots,K\}.
\end{equation}
Then \(P_d(\theta_0)-P_d(\theta_{j,-})=\Delta_j\) and \(P_d(\theta_{j,+})-P_d(\theta_0)=\Delta_j\).
Both parameters lie in \([1/3,2/3]\), uniformly for \(d\ge2\).
Indeed, \(1-1/d\ge1/2\) gives \(\theta_{j,-}^2\ge3/16>1/9\) and \(\theta_{j,+}^2\le5/16<4/9\).
On this interval,
\[
\frac13\le\odv{P_d}{\theta}(\theta)=2(1-1/d)\theta\le\frac43.
\]
The mean value theorem applied to \(P_d(\theta_{j,+})-P_d(\theta_{j,-})=2\Delta_j\) therefore gives
\begin{equation}
\label{eq:id-parameter-gap}
\frac32\Delta_j\le h_j:=\theta_{j,+}-\theta_{j,-}\le6\Delta_j.
\end{equation}

Fix a unit vector \(\vec e_0\), draw \(\vec\psi_2,\ldots,\vec\psi_K\) independently from Haar measure once, and define the base prior
\[
\nu=(\vec\rho_{\theta_0,\vec e_0},\vec\rho_{\theta_{2,-},\vec\psi_2},\ldots,\vec\rho_{\theta_{K,-},\vec\psi_K}),\qquad \Pi_0=\operatorname{Law}(\nu).
\]
For each challenger \(j\), change its parameter and retain its direction:
\[
\vec\rho_i^{(j)}=\begin{cases}
\vec\rho_{\theta_{j,+},\vec\psi_j},&i=j,\\
\vec\rho_i,&i\ne j,
\end{cases}
\qquad \nu^{(j)}=(\vec\rho_i^{(j)})_{i=1}^K.
\]
Writing \(\lambda_i^{(j)}=P(\vec\rho_i^{(j)})\), the base and alternative gaps satisfy
\begin{align*}
\lambda_1-\lambda_i&=\Delta_i&&\text{for }i\ge2,\\
\lambda_j^{(j)}-\lambda_1^{(j)}&=\Delta_j,\\
\lambda_j^{(j)}-\lambda_i^{(j)}&=\Delta_j+\Delta_i&&\text{for }i\notin\{1,j\}.
\end{align*}
Thus every base instance and every alternative belongs to \(\mathcal V_{\underline\Delta}\).
This common prior depends on the prescribed gaps and is independent of the algorithm.
For events \(B\) and nonnegative random variables \(X\), write
\[
\Pr_{\Pi_0}(B)=\int\Pr_\nu(B)\,\odif{\Pi_0}(\nu),\qquad \E_{\Pi_0}X=\int\E_\nu X\,\odif{\Pi_0}(\nu).
\]
Let \(\tau\) be the terminal round of the original algorithm, and let \(N_j(T)\) count all copies of source \(j\) requested along its run, including a run that continues indefinitely.
The convention \(T=+\infty\) on a nonterminating run gives \(\sum_jN_j(T)\le T\).
On every base instance, the event \(\mathcal C:=\{T<\infty,\widehat i=1\}\) has probability at least \(1-\delta\).

The confidence term admits a sourcewise probability bound on every deterministic base instance.
Fix such a \(\nu\) and a challenger \(j\).
The changed states commute and share their distinguished eigenvector.
Set \(u_{j,\pm}=\theta_{j,\pm}+(1-\theta_{j,\pm})/d\), their eigenvalues on that vector.
Each of their remaining eigenvalues equals \((1-u_{j,\pm})/(d-1)\).
Consequently, for a universal \(C_{\mathrm D}>0\),
\begin{align*}
D(\vec\rho_{\theta_{j,-},\vec\psi_j}\Vert\vec\rho_{\theta_{j,+},\vec\psi_j})
&=\kl(u_{j,-},u_{j,+})\\
&\le\frac{(u_{j,-}-u_{j,+})^2}{u_{j,+}(1-u_{j,+})}\le C_{\mathrm D}\Delta_j^2.
\end{align*}
The first inequality applies \(\log x\le x-1\) to both terms, whose resulting sum is \(p(p-q)/q+(1-p)(q-p)/(1-q)=(p-q)^2/[q(1-q)]\).
The last inequality is uniform because \(u_{j,\pm}\in[1/3,5/6]\) and \(\lvert u_{j,+}-u_{j,-}\rvert\le h_j\le6\Delta_j\).
Put \(A_j:=\Delta_j^{-2}\log(1/\delta)\), \(a_{\mathrm s}:=(16C_{\mathrm D})^{-1}\), and \(m:=\lfloor a_{\mathrm s}A_j\rfloor\).
Suppose that \(\Pr_\nu(N_j(T)>m)\le1/2\).
Then
\[
\Pr_\nu(\mathcal C\cap\{N_j(T)\le m\})\ge\frac12-\delta\ge\frac7{16}.
\]
The events \(\mathcal C\cap\{N_j(T)\le m,\tau\le H\}\) increase to this event as \(H\to\infty\).
Choose a finite integer \(H\) such that their probability is at least \(3/8\).
Run the original algorithm with two caps: at most \(H\) rounds and at most \(m\) copies from source \(j\).
Process a normal stopping decision before either cap, and stop before any request that would exceed the source cap.
Return a binary output \(B=1\) if the algorithm returns source \(1\), and return \(B=0\) for every other answer or cap.
The caps are the same under \(\nu\) and \(\nu^{(j)}\).
This finite experiment satisfies
\[
p:=\Pr_\nu(B=1)\ge\frac38,\qquad q:=\Pr_{\nu^{(j)}}(B=1)\le\delta.
\]
Applying Lemma~\ref{lem:id-transportation} to this capped experiment gives
\begin{equation}
\label{eq:id-confidence-charge}
\kl(p,q)\le mD(\vec\rho_{\theta_{j,-},\vec\psi_j}\Vert\vec\rho_{\theta_{j,+},\vec\psi_j})\le mC_{\mathrm D}\Delta_j^2\le\frac1{16}\log\frac1\delta.
\end{equation}
Binary entropy is at most \(\log2\), so
\[
\kl(p,q)\ge p\log\frac1q-\log2\ge\frac38\log\frac1\delta-\log2\ge\frac18\log\frac1\delta,
\]
since \(\delta\le1/16\).
The case \(q=0\) follows with infinite relative entropy.
This contradicts Equation~\eqref{eq:id-confidence-charge}.
Because \(N_j(T)\) is integer valued, it follows that
\begin{equation}
\label{eq:id-pointwise-confidence}
\Pr_\nu(N_j(T)>a_{\mathrm s}A_j)>\frac12.
\end{equation}

For the memory term, consider an unknown source \(\vec\rho_{\theta,\vec\psi}\), with its Haar direction drawn once and \(\theta\in\{\theta_{j,-},\theta_{j,+}\}\), and use it as source \(j\) in a simulation of the identification algorithm.
Independently sample every other challenger's direction according to \(\Pi_0\) and keep its classical description.
The reference state has a known classical description as well.
For a requested pair from auxiliary source \(i\), let \(\vec F_h(B)\) be the selected two-stage measurement effect for a measurable set \(B\) of complete classical records, conditional on the current history \(h\).
The simulator samples from the kernel
\[
\mathsf K_i(h,B)=\tr(\vec F_h(B)\vec\rho_i^{\otimes2}).
\]
The auxiliary state's description is fixed throughout the simulation, and this kernel is parameter independent once that description and the history are fixed.
It preserves every outcome used to select later operations.
Resetting quantum memory after each pair makes this classical simulation exact across rounds.
With unrestricted classical computation, standard Borel kernels admit exact measurable sampling using independent uniform seeds.
These auxiliary seeds and descriptions are independent of both the unknown parameter and its hidden direction.
The base law of the simulation is the original interaction averaged over \(\Pi_0\), and the alternative law is the corresponding mixture of \(\nu^{(j)}\).

Suppose that \(\Pr_{\Pi_0}(N_j(T)>m)\le1/2\) for an integer \(m\ge0\).
The same increasing-event argument gives a finite simulation horizon \(H\) such that the source cap and horizon cap produce a binary output \(B\) with
\[
\Pr_{\theta_{j,-}}(B=1)\ge\frac38,\qquad \Pr_{\theta_{j,+}}(B=1)\le\delta.
\]
Here a simulation step is one complete pair from source \(j\) or one auxiliary classical transition, and the output and cap rules are the same as above.
The simulation uses at most \(J:=\lfloor m/2\rfloor\) pairs from the tested source.
After it returns, append discarded fresh pairs to reach exactly \(J\) rounds, preserving \(B\).
The finite horizon allows all auxiliary transitions before the next tested pair to be composed into one classical transition kernel.
Its dependence on \(\theta,\vec\psi\) is entirely through the preceding history.
Include the initial auxiliary descriptions, seeds, and classical records in the complete transcript.
The resulting fixed-budget experiment is in the reset model with retained dimension at most \(r\).
If \(J=0\), its law is parameter independent, contradicting the difference of at least \(3/8-\delta\ge5/16\) between the two output probabilities.
For \(J\ge1\), write \(Q_\theta\) for its complete record law.
Lemma~\ref{lem:information}, Equation~\eqref{eq:testing-main}, and output data processing give
\begin{align}
\frac5{16}
&\le\TV(P_{\theta_{j,-}}^B,P_{\theta_{j,+}}^B)
\le\TV(Q_{\theta_{j,-}},Q_{\theta_{j,+}})\nonumber\\
&\le\frac12\int_{\theta_{j,-}}^{\theta_{j,+}}\sqrt{I(Q_\theta)}\,\odif{\theta}
\le C h_j\sqrt{\min\ab\{J,\frac{J^2+Jr}{d}\}}.
\label{eq:id-capped-information}
\end{align}
The common reference densities and derivative bounds from Appendix~\ref{sec:regularity} apply because every auxiliary transition is parameter independent conditional on its history and initial seed.
Equation~\eqref{eq:id-capped-information} implies universal \(b_0,b_1>0\) such that
\[
J\ge b_0h_j^{-2},\qquad J^2+Jr\ge b_1d/h_j^2.
\]
The second inequality forces either \(J^2\ge b_1d/(2h_j^2)\) or \(Jr\ge b_1d/(2h_j^2)\).
Combining these two possibilities with the first inequality gives a universal \(b>0\) such that
\[
J\ge b\Rrate{h_j}\ge\frac b{36}\Rrate{\Delta_j}.
\]
The final inequality uses \(h_j\le6\Delta_j\), which decreases each inverse-square term by at most a factor of \(36\) and the inverse-linear term by at most a factor of \(6\).
Set \(a_{\mathrm m}:=b/36\) and \(m:=\lfloor a_{\mathrm m}\Rrate{\Delta_j}\rfloor\).
Then \(J\le m/2\le(b/72)\Rrate{\Delta_j}\), contradicting the preceding lower bound.
Thus
\begin{equation}
\label{eq:id-memory-tail}
\Pr_{\Pi_0}(N_j(T)>a_{\mathrm m}\Rrate{\Delta_j})>\frac12.
\end{equation}

Let
\[
S_j:=\max\ab\{\Delta_j^{-2}\log(1/\delta),L_{d,r}(\Delta_j)\},\qquad a:=\min\{a_{\mathrm s},a_{\mathrm m}\}.
\]
Since \(\log(1/\delta)\ge1\), we have \(S_j=\max\{A_j,\Rrate{\Delta_j}\}\).
Average Equation~\eqref{eq:id-pointwise-confidence} over \(\Pi_0\) and combine it with Equation~\eqref{eq:id-memory-tail}.
For each source, the bound corresponding to the larger of \(A_j\) and \(\Rrate{\Delta_j}\) gives
\begin{equation}
\label{eq:id-prior-sourcewise}
\Pr_{\Pi_0}(N_j(T)>aS_j)\ge\frac12,\qquad j\in\{2,\ldots,K\}.
\end{equation}
Define the bounded random variable and its deterministic upper bound
\[
Y:=\sum_{j=2}^K aS_j\ind\{N_j(T)>aS_j\},\qquad W:=a\sum_{j=2}^K S_j.
\]
Then \(0\le Y\le W\), \(Y\le T\), and Equation~\eqref{eq:id-prior-sourcewise} gives \(\E_{\Pi_0}Y\ge W/2\).
Writing \(p_*:=\Pr_{\Pi_0}(Y\ge W/4)\), boundedness gives
\[
\frac W2\le\E_{\Pi_0}Y\le\frac W4(1-p_*)+Wp_*,
\qquad p_*\ge\frac13.
\]
Consequently
\[
\Pr_{\Pi_0}(\mathcal C\cap\{T\ge W/4\})\ge\frac13-\delta\ge\frac{13}{48}>\frac14.
\]
There is therefore a deterministic base instance \(\nu\) with the prescribed gaps such that
\[
\Pr_\nu\ab(T<\infty,\widehat i=1,T\ge\frac a4\sum_{j=2}^K S_j)\ge\frac14.
\]
Taking \(c=a/4\) proves Equation~\eqref{eq:id-lower}.
For equal gaps, the family lies in \(\mathcal V_\Delta\) and the sum is \((K-1)\max\{\Delta^{-2}\log(1/\delta),L_{d,r}(\Delta)\}\).
Since \(\delta\le1/16<1/4\), any copy budget attaining probability at least \(1-\delta\) for correct identification within that budget must be at least the stated lower-bound scale.
\end{proof}

\end{document}